\documentclass[pra,aps,nofootinbib,notitlepage,twocolumn,superscriptaddress]{revtex4-1}

\usepackage{amsmath}
\usepackage{amssymb}
\usepackage{graphicx}

\usepackage{txfonts}
\usepackage{dsfont}

\usepackage[colorlinks=true,linkcolor=blue,citecolor=blue,urlcolor=blue]{hyperref}

\providecommand{\abs}[1]{\left\lvert#1\right\rvert}
\providecommand{\norm}[1]{\lVert#1\rVert}

\providecommand{\tr}{{\rm tr}}
\renewcommand{\phi}{\varphi}

\begin{document}
\title{Mode-Dependent Loss Model for Multimode Photon-Subtracted States}

\author{Mattia Walschaers} 
\email{mattia.walschaers@lkb.upmc.fr}
\affiliation{Laboratoire Kastler Brossel, Sorbonne Universit\'e, ENS-PSL Research University, Coll\`ege de France, CNRS; 4 place Jussieu, F-75252 Paris, France}
\author{Young-Sik Ra}
\affiliation{Laboratoire Kastler Brossel, Sorbonne Universit\'e, ENS-PSL Research University, Coll\`ege de France, CNRS; 4 place Jussieu, F-75252 Paris, France}
\affiliation{Department of Physics, Korea Advanced Institute of Science and Technology (KAIST), Daejeon 34141, Korea}
\author{Nicolas Treps}
\affiliation{Laboratoire Kastler Brossel, Sorbonne Universit\'e, ENS-PSL Research University, Coll\`ege de France, CNRS; 4 place Jussieu, F-75252 Paris, France}

\date{\today}

\begin{abstract}
Multimode photon-subtraction provides an experimentally feasible option to construct large non-Gaussian quantum states in continuous-variable quantum optics. The non-Gaussian features of the state can lead towards the more exotic aspects of quantum theory, such as negativity of the Wigner function. However, the pay-off for states with such delicate quantum properties is their sensitivity to decoherence. In this paper, we present a general model that treats the most important source of decoherence in a purely optical setting: losses. We use the framework of open quantum systems and master equations to describe losses in $n$-photon-subtracted multimode states, where each photon can be subtracted in an arbitrary mode. As a main result, we find that mode-dependent losses and photon-subtraction generally do not commute. In particular, the losses do not only reduce the purity of the state, they also change the modal structure of its non-Gaussian features. We then conduct a detailed study of single-photon subtraction from a multimode Gaussian state, which is a setting that lies within the reach of present-day experiments.
\end{abstract}

\maketitle

\section{Introduction}

In a time where quantum technologies are gradually becoming a reality, the attention for potential physical implementations of quantum computers increases. An important open question deals with the platform on which these quantum information processors will ultimately be developed. Serious contenders include solid-state architectures in semi- and superconductors \cite{veldhorst_two-qubit_2015,Watson:2018aa,Neill195,Rigetti-2012}, nitrogen vacancy centres in diamond \cite{Wrachtrup:2006aa,Hensen:2015aa}, trapped ions \cite{Monz:2016aa,Schafer:2018aa}, and light \cite{obrien_photonic_2009}. In general, quantum properties within these systems are suppressed by interactions to an uncontrollable environment, which induces decoherence. Light's resilience against such detrimental decoherence effects thus offers an advantage when it is used to process quantum information.

Setups that rely strongly on the use and manipulation of individual photons are confronted with another difficulty: the controlled generation of sufficiently large numbers of photons \cite{Wang:2017aa} and the number-resolved detection \cite{humphreys_tomography_2015} thereof. This ultimately implies that it is hard to scale photonic quantum devices. Therefore, one can alternatively resort to treating light in the continuous variable (CV) regime. This implies that the observables of interest are the field quadratures, i.e.~the real and imaginary parts of the complex amplitude of the electromagnetic wave. Large entangled states can be deterministically generated  in this setting \cite{cai-2017,gerke_full_2015,PhysRevLett.112.120505,roslund_wavelength-multiplexed_2014,Su:12,yoshikawa_invited_2016}, which can be used for measurement-based quantum computation \cite{gu_quantum_2009}.

A crucial ingredient in universal CV quantum computation is the ability to induce non-Gaussian statistics for quadrature measurements. This turns out to be a challenging task from the experimental point of view \cite{arzani_polynomial_2017,PhysRevA.84.053802,PhysRevA.93.022301}. An experimentally feasible way to achieve this goal, is through photon subtraction \cite{dakna_generating_1997,parigi_probing_2007,zavatta_experimental_2007,zavatta_subtracting_2008}. This method can also be generalised to light with many optical modes \cite{averchenko_multimode_2016,PhysRevA.89.063808,ra_tomography_2017,Ra2019,walschaers_statistical_2017,walschaers_entanglement_2017}, where it can also enhance entanglement between modes \cite{PhysRevA.86.012328,ourjoumtsev_increasing_2007,takahashi_entanglement_2010,walschaers_entanglement_2017}.

Even though such photon-subtracted states of light may hold advantages for quantum information processing, such as scalability and resilience to noise, there are also barriers along the way. In order to produce and manipulate the light, one uses a wide range of linear and non-linear optical elements that can have unwanted side effects that lead to a decrease of quantum properties. The most notable of these effects is photon loss, which can arise in a variety of ways. A notable way of modelling such losses involves the physics of open quantum systems \cite{alicki_quantum_1987,alicki_theory_1978,davies_irreversible_1977,gorini_completely_1976,lindblad_generators_1976,gardiner-book,breuer_theory_2007}. This approach was successfully applied to study the effect of losses on the single-mode photon-subtracted vacuum state \cite{biswas_nonclassicality_2007}. In this paper, we will use the open systems approach to develop a general loss model for multimode states with an arbitrary number of photon subtractions in arbitrary modes, hence generalising the result of \cite{biswas_nonclassicality_2007}. We will also show that this result generalises the scenario where losses are modelled through beamsplitters. 

We start, in Section \ref{sec:multimode}, by introducing the subtleties of multimode quantum optics and fixing our notation. The open system loss model is introduced in Section \ref{sec:open}, where we show its equivalence to the beamsplitter model in certain scenarios. Our main result, describing losses in arbitrary multimode photon-subtracted states is presented in Section \ref{sec:result}. This result is then detailed in the specific context of single-photon subtraction in Section \ref{sec:result1photon}, where it leads to a specific and simple modification of the Wigner function of \cite{walschaers_entanglement_2017}. In the latter section we also illustrate our results with some examples.

\section{Multimode quantum optics}\label{sec:multimode}

We first introduce the framework of continuous variables in multimode optical systems. CV quantum optics relies on {\em quadratures} of the electromagnetic field as the relevant quantum observables. In multimode quantum optics, the electric field operator $\hat{E}(\mathbf{r},t)$ is expressed in terms of a basis $\{u_1({\bf r}, t), \dots, u_m({\bf r},t)\}$ of $m$ normalised modes and their associated {\em amplitude and phase quadrature operators}, $\hat{x}_j$ and $\hat{p}_j$, respectively: \footnote{These $u_j({\bf r}, t)$ are solutions to Maxwell's equations, normalised with respect to the spacial degrees of freedom, i.e.~$\frac{1}{V}\int {\rm d}^3 r\, \abs{u_j({\bf r},t)}^2  = 1$ for every time $t$.}
\begin{equation}
\hat{E}(\mathbf{r},t) = \epsilon_c\sum_{j=1}^m (\hat{x}_j + i \hat{p}_j)u_j({\bf r}, t),
\end{equation}
where $\epsilon_c$ is a constant that carries the dimension of the field.
Moreover, because these modes energetically behave as harmonic oscillators, the quadratures follow the canonical commutation relations $[\hat{x}_j, \hat{p}_k] = 2i \delta_{j,k}$, $[\hat{x}_j, \hat{x}_k] = 0$, and $[\hat{p}_j, \hat{p}_k] =0$. 

In our present work, it is convenient to define a general quadrature operator $Q(f)$ as
\begin{equation}\label{eq:Qgen}
Q(f) \equiv \sum_{k=1}^m (f_k \hat{x}_k + f_{k + m} \hat{p}_k),
\end{equation}
where $f \in{\cal N}(\mathbb{R}^{2m})$, with ${\cal N}(\mathbb{R}^{2m})$ the set of normalised vectors in the {\em optical phase space} $\mathbb{R}^{2m}$. In addition, the optical phase space is equipped with a symplectic structure that connects amplitude and phase quadratures of the same mode. This symplectic structure can be represented by a matrix $J$ that acts on the phase space, with $J^2= -\mathds{1}$ and $J^T = -J$. With this symplectic structure, we can generalise the canonical commutation relation to
\begin{equation}
[Q(f_1),Q(f_2)] = -2i(f_1, Jf_2),\, \text{ for all $f_1,f_2 \in{\cal N}(\mathbb{R}^{2m})$,}
\end{equation}
where $(.,.)$ denotes the inner product on $\mathbb{R}^{2m}$.
Moreover, this allows us to define general creation and annihilation operators as
\begin{equation}\label{eq:crean}
a^{\dag}(f) = \frac{1}{2}[Q(f) - iQ(Jf)],\, \text{ and } a(f) = \frac{1}{2}[Q(f) + iQ(Jf)].
\end{equation}
Note that the symplectic transformation, induced by $J$, causes a $\pi/2$ phase shift, i.e.~$a^{\dag}(Jf) = i a^{\dag}(f)$. These operators play a crucial role in describing loss processes in quantum optics.

\section{Open System Model}\label{sec:open}

The framework of open quantum systems is ubiquitous in quantum physics, as it describes how a small (typically controllable) quantum system is embedded in a large (typically uncontrollable environment) \cite{breuer_theory_2007,gardiner-book}. A common approach to such systems uses a master equation that describe a non-unitary evolution. This can ultimately capture a wide range of phenomena, where concepts such as (non-)Gaussianity, correlation, (non-)Markovianity, et cetera play an important role. Throughout the following section, we gradually build our specific noise model by adding a range of assumptions that will ultimately lead us to an analytically tractable -- though realistic -- formalism.

\subsection{Completely Positive Maps and the Master Equation}

In general, we describe the effect of losses (or any coupling to an environment) in the {\em Heisenberg picture} by a channel $\Lambda: {\cal A} \mapsto {\cal A}$, where ${\cal A}$ is the algebra of observables. In our specific case, ${\cal A}$ is generated by the quadrature operators $Q(f)$. The channel $\Lambda$ has to fulfil the following basic criteria:
\begin{align}
&\Lambda(\mathds{1}) = \mathds{1},\\
&\Lambda(x^{\dag}x) \geqslant 0, \quad \text{ for all $x \in {\cal A}$}.\label{eq:pos}\\
&\text{$\Lambda$ is linear.}
\end{align}
It is common to strengthen (\ref{eq:pos}) by adding 
%an $N$-dimensional ancilla, to obtain ${\cal A} \otimes {\cal M}(N)$,\footnote{Note that ${\cal M}(N)$ denotes the $N$-dimensional matrix algebra} and trivially extending the channel as $\Lambda \otimes {\rm id}$.\footnote{Here, ${\rm id}$ denotes the identity operation such that ${\rm id}(x) = x$ for all $x \in {\cal A}$.} The then demand that $\Lambda \otimes {\rm id}(y^{\dag}y) \geqslant 0$ for all $y \in {\cal A} \otimes {\cal M}(N)$. If this holds for all possible dimensions $N$ for the ancilla, we say
the demand that $\Lambda$ is {\em completely positive} \cite{stinespring_positive_1955}, i.e.~that the channel can be represented by a Kraus representation.
%acts correctly when there are certain additional degrees of freedom in the system that are not being considered explicitly.

In order to describe our loss model, we will assume that this channel depends on an overall parameter $\xi \geqslant 0$ that characterises the strength of the losses. Note that this does not imply that every mode in the system has the same losses, $\xi$ simply acts as an overall scaling factor. It is common in open system models that this parameter represent the evolution time of the system (long evolution times typically imply high losses). However, in optics time acts in a very different way than in mechanical systems, and therefore we will simply consider $\xi$ as a parameter.\footnote{Note that this is similar to cases where we model optical elements through a Hamiltonian. When these Hamiltonians are exponentiated to obtain the associated unitary transformation, time is also replaced by a more generic parameter.} We now make the additional demands on $\Lambda_{\xi}$ that
\begin{align}
&\Lambda_{\xi = 0}(x)= x, \quad \text{ for all $x \in {\cal A}$,}\\
&\Lambda_{\xi} \circ \Lambda_{\zeta} = \Lambda_{\xi + \zeta}.\quad  \text{ for all $\xi, \zeta \geqslant 0$.} \label{eq:comp}
\end{align}
Because of these properties, the channel is said to be a {\em one-parameter semigroup}, which implies that there is a composition rule for channels (\ref{eq:comp}). Note, however, that induced noise can typically not be undone, since $\Lambda_{\xi}$ does not necessarily have an inverse operation. Because the composition rule (\ref{eq:comp}) holds for all parameters $\xi$ and $\zeta$, this channel is a {\em Markovian} map  \cite{breuer_theory_2007,gardiner-book}. 
%It follows that there is no notion of memory in the channel as we can divide it in arbitrary many subsequent ``sub-channels''.

An important theorem for the generation of such completely positive semigroups was presented in \cite{gorini_completely_1976,lindblad_generators_1976}. It was shown that completely positive semigroup $\Lambda_{\xi}$ can be generated through a differential equation of the form
\begin{equation}\label{eq:lindblad}
\frac{\rm d}{{\rm d} \xi} \Lambda_{\xi}(x) \equiv i [H, x] + \sum_j \left( l_j^{\dag} x l_j -\frac{1}{2} \{l_j^{\dag} l_j , x\}\right), \quad {x \in {\cal A}},
\end{equation}
where $\{.,.\}$ denotes the anti-commutator. The operator $H = H^{\dag}  \in {\cal A}$ is the system's Hamiltonian, and $l_j  \in {\cal A}$ are the Lindblad operators. These Lindblad operators typically describe the interactions between the system and its environment. The Hamiltonian, on the other hand, describes unitary transformations on the system; it could, for example, be used to include linear optics in a model. However, we intend to develop a loss model, and, therefore, we set $H=0$ in the remainder of the paper.\\

To ultimately derive the loss model for photon-subtracted states, we will not only restrict ourselves to the Heisenberg picture. It turns out that it is convenient to interchange between both the Heisenberg and the Schr\"odinger picture. The latter in particular allows us to understand the effect of losses on the level of the Wigner functions. Hence, we define $\Lambda^{\star}$ as the Schr\"odinger picture equivalent of $\Lambda$, which is contained in the identity
\begin{equation}\label{eq:identity}
\tr [ \rho \Lambda(x) ] = \tr [\Lambda^{\star}(\rho)x], 
\end{equation}
which holds for all observables $x \in {\cal A}$ and all states $\rho$.\\

%Formally, there are a range of mathematical subtleties that must be taken into account when we are dealing with bosonic systems. They fall beyond the scope of the present work, but are discussed in detail in Chapter 9 of \cite{walschaers_efficient_2016}. 
In the following section, we will identify the specific choice of Lindblad operators that must be inserted in (\ref{eq:lindblad}) to obtain our loss model.

\subsection{Losses in Optical Systems}\label{sec:lossopen}

General loss models have been considered in a wide range of literature \cite{PhysRevA.31.1059,CELEGHINI1992156,PhysRevA.75.013811,Vogel-Welsch,Agarwal-Book}. 
%As a rule of thumb, the Lindblad operators $l_i$ in (\ref{eq:lindblad}) have some connection to annihilation operators in order to introduce the loss. However, there is still a range of possibilities. For example, losses can be correlated, such that groups of multiple photons can disappear together.\footnote{This can be achieved by setting 
%for example $l_j = a(f_{j})a(f_{j+1})$, where $f_{j}, f_{j+1} \in{\cal N}(\mathbb{R}^{2m})$ denoted arbitrary modes.}
In the present model, we will assume that the loss process is Gaussian, i.e.~that Gaussian states are mapped into Gaussian states \cite{demoen_completely_1977,demoen_completely_1979, alicki_quantum_1987,alicki_theory_1978}. Hence, we set the Lindblad operators 
\begin{equation}l_j = \sqrt{\gamma_j}\,a(h_j), \quad h_{j} \in{\cal N}(\mathbb{R}^{2m}), \label{eq:lindbladops}\end{equation}
where $\gamma_j$ denotes the loss parameter of the mode $h_j$, that multiplies the overall strength of the losses $\xi$.

Our method to analytically solve equation (\ref{eq:lindblad}), and obtain the loss channel $\Lambda_{\xi}$, is based on earlier work \cite{davies_irreversible_1977} that was explicitly adapted for the bosonic case in \cite{Walschaers2018,walschaers_optimal_2017}. There, the general result for the action of $\Lambda_{\xi}$ on a normally ordered monomial of creation and annihilation operators is given:
\begin{equation}\begin{split}\label{eq:lambdaHeisSol}
\Lambda_{\xi} &[a^{\dag}(f_1)\dots a^{\dag}(f_r)a(f_{r+1})\dots a(f_s)] \\ &= a^{\dag}(e^{-\xi D}f_1)\dots a^{\dag}(e^{-\xi D}f_r)a(e^{-\xi D}f_{r+1})\dots a(e^{-\xi D}f_s).
\end{split}
\end{equation}
where
\begin{equation}\label{eq:D}
D = \sum_{j=1}^m \frac{\gamma_j}{2} (P_{h_j}+P_{Jh_j}).
\end{equation}
Here $P_{h_j} $ is a projection operator on the phase space vector $h_j\in{\cal N}(\mathbb{R}^{2m})$. The formal definition of creation and annihilation operators with non-normalised vectors is provided in Appendix \ref{app:creation}. Note that $0 \leqslant \exp(-\xi D) \leqslant \mathds{1}$ for all possible $\xi \geqslant 0$. We can then also find the natural property that for any $f\in{\cal N}(\mathbb{R}^{2m})$ \begin{equation}
\xi > \xi' \implies \Lambda_{\xi}[a^{\dag}(f)a(f)] \leqslant  \Lambda_{\xi'}[a^{\dag}(f)a(f)],
\end{equation}
such that the number of particles decays with increasing values of $\xi$.\\

To conclude our discussion on the open-system approach to losses, we show its effect on a Gaussian state $\rho_G$. Such a state is fully characterised by its first and second order quadrature correlations, $\tr [\rho_G Q(f)]$ and $\tr [\rho_G Q(f_1)Q(f_2)]$, respectively. By using that $Q(f) = a(f) +a^{\dag}(f)$, we find that 
\begin{align}
&\Lambda_{\xi}[Q(f)] = Q(e^{-\xi D}f),\\
&\Lambda_{\xi}[Q(f_1)Q(f_2)] = Q(e^{-\xi D}f_1)Q(e^{-\xi D}f_2) + (\mathds{1} - e^{-2 \xi D}).
\end{align}
We then find that the action of the loss channel on the state's covariance matrix $V$ is given by
\begin{equation}\label{eq:Vxi}
V \overset{\Lambda_{\xi}}{\mapsto} e^{-\xi D}Ve^{-\xi D} + (\mathds{1} - e^{-2 \xi D}) \equiv V_{\xi}.
\end{equation}
When we then assume uniform losses (i.e.~losses that are the same in every mode) and set $D = \mathds{1}/2$, we find that $V \mapsto e^{-\xi}V + (1 - e^{-\xi}) \mathds{1}.$ In other words, we are mixing a Gaussian state with the vacuum. We will see that such a result is also obtained when losses are modelled through a beamsplitter.

\subsection{Equivalence to Beamsplitter Model}\label{sec:BeamsplitterModel}

The model presented in Section \ref{sec:lossopen} can be equivalently represented by means of beamsplitters. We start this discussion with the single-mode case. A beamsplitter mixes our mode of interest with an auxiliary mode which is in a vacuum state. We can describe this mixing of modes by a unitary transformation \begin{equation}
U = \begin{pmatrix}t & -r\\ r & t\end{pmatrix},
\end{equation}
where we choose $r, t \in \mathbb{R}$ for simplicity.\footnote{Generally, the entries of a beamsplitter can be complex, such that it also changes the phase of the mode. To include such an effect in the open system model, one must set $h \neq 0$ in (\ref{eq:lindblad}).} We then demand that $t^2 + r^2 = 1$ to guarantee unitarity of the operation. With this step we can mix the creation or annihilation operators ($a^{\dag}$ and $a$, respectively) for the mode of interest with the creation or annihilation operators ($b^{\dag}$ and $b$, respectively) of the auxiliary mode. As such, the beamsplitter carries out the mapping
\begin{align}
& a^{\dag} \overset{U}{\mapsto} t a^{\dag} + r b^{\dag},\\
& a  \overset{U}{\mapsto} t a + r b.
\end{align}
However, the auxiliary mode is assumed to be in a vacuum state, which extends the quantum state of the system to $\rho \otimes \lvert  0\rangle \langle 0\rvert$, with $\rho$ an arbitrary state for the mode of interest. Intuitively, this implies that we must act with the vacuum state on the creation and annihilation operators associated with the auxiliary mode.
%To make that statement rigorous, we introduce the algebras of observables for the mode of interest and for the auxiliary mode, ${\cal A}^{(1)}$ and ${\cal B}^{(1)}$, respectively.\footnote{The superscript $(1)$ is introduced to indicate that we are just working with a single mode at this point.} We can map the problem back to a single-mode problem in the Heisenberg picture by executing the map $\Pi: {\cal A}^{(1)}\otimes {\cal B}^{(1)} \rightarrow {\cal A}^{(1)}$, given by $\Pi = {\rm id} \otimes \omega_0$ with $\omega_0(x) = \langle 0\rvert x \lvert  0\rangle$ for all $x \in {\cal B}^{(1)}$. In particular, this means that the map acts as\footnote{Note that in explicit tensor product notation $a^{\dag}$ is actually $a^{\dag} \otimes \mathds{1}$ and $b^{\dag}$ is $\mathds{1}\otimes b^{\dag}$.}
%\begin{align}
%\Pi(T a^{\dag} + R b^{\dag}) = T a^{\dag}
%\end{align}
%The action of the loss channel $\Lambda$, modelled through a beamsplitter model, is obtained by assuming that the auxiliary mode is prepared in the vacuum state, which is particularly convenient for normally ordered products of creation and annihilation operators. 
As such, we obtain the action of $\Lambda$ on normally ordered products of creation and annihilation operators
%we obtain 
%\begin{equation}
%\Lambda(a^{\dag}) = T a^{\dag}, \, \text{ and } \Lambda(a) = T a.
%\end{equation}
%We can now apply this operation to 
\begin{equation}
\Lambda [(a^{\dag})^r a^{s-r}] = t^{s}(a^{\dag})^r a^{s-r}.
\end{equation}
This result is in perfect agreement with the single-mode version of (\ref{eq:lambdaHeisSol}), with $t = \exp(-\xi\gamma/2)$.

The beamsplitter result can trivially be extended to a multimode beamsplitter, which has the same transmittance and reflectivity for every mode. In this case, we find that 
\begin{equation}\begin{split}
\Lambda &[a^{\dag}(f_1)\dots a^{\dag}(f_r)a(f_{r+1})\dots a(f_s)] \label{eq:beamsplitterManyMode} \\
&= t^{s} a^{\dag}(f_1)\dots a^{\dag}(f_r)a(f_{r+1})\dots a(f_s).
\end{split}
\end{equation}
This result is also compatible with (\ref{eq:lambdaHeisSol}), with $t = \exp(-\xi\gamma/2)$, by setting $D = \gamma \mathds{1}/2$.\\

\begin{figure}
\centering
\includegraphics[width=0.35\textwidth]{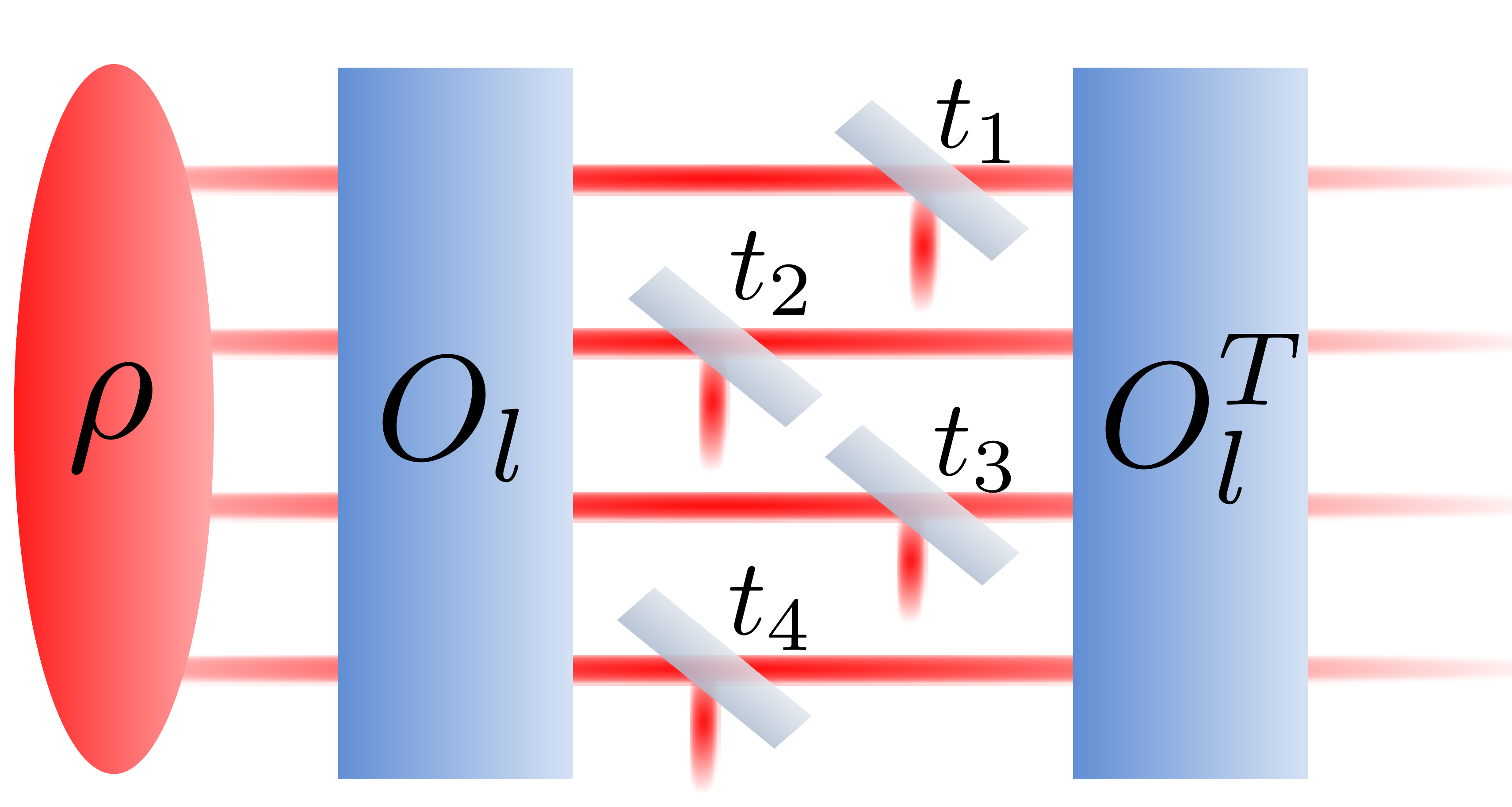}
\caption{Schematic representation of mode-dependent losses in a multimode setting, using the beamsplitter representation. The basis change (i.e.~linear interferometer) $O_l$ transforms the mode basis in which the state $\rho$ is given to the mode basis $\{h_1, \dots, h_4\}$ in which the losses act locally. As described in the main text, in this basis the losses can be described by a beamsplitter model with transmittance $t_j$ for mode $h_j$. \label{fig:LossesIllust}}
\end{figure}

Full equivalence between both models is obtained by choosing a basis of modes and selecting a distinct beamsplitter for each one of the modes. One can then tune the transmittance of each one of these beamsplitters at will to obtain the result (\ref{eq:lambdaHeisSol}). In particular, one must choose the basis of modes to be $\{h_1, \dots, h_m\}$ and set the transmittance $t_j = \exp(-\xi \gamma_j/2)$. We will not go through the detailed derivation for this multimode scenario, but rather present a schematic representation of concept in Fig.~\ref{fig:LossesIllust}. In this sketch, we assume that the state $\rho$ is represented in a fixed mode basis, whereas the losses occur in local modes $\{h_1, \dots, h_4\}$. The orthogonal symplectic matrix $O_l$ represents a mode basis change (i.e. a linear interferometer) that maps the modes in which $\rho$ is given to the mode basis $\{h_1, \dots, h_4\}$. The final addition of a second basis change, given by $O_l^T$ serves to recast the state in the original mode basis of the state $\rho$.

\section{Multimode Photon-Subtracted States}\label{sec:result}

\subsection{Algebraic approach}\label{sec:AlgApproach}

In the previous sections, we explained how the loss channel's action on any normally ordered monomial is described by (\ref{eq:lambdaHeisSol}). In this section, we will use this result to explain the effect of the loss channel on a photon-subtracted state. We will use the result in the Heisenberg picture, to obtain the associated map in the Schr\"odinger picture.\\

We start by considering an arbitrary state $\rho$ of the multimode optical system under consideration. We can then subtract photons from this state by acting on it with annihilation operators. In the most general case, we can subtract $n$ photons from a set of modes $g_1, \dots, g_n \in{\cal N}(\mathbb{R}^{2m})$. Note that we do not require these modes to be orthogonal. On a formal level, the photon-subtracted state is described by
\begin{equation}
\rho_{-} = \frac{a(g_1)\dots a(g_n) \rho a^{\dag}(g_n) \dots a^{\dag}(g_1) }{{\rm tr}[a^{\dag}(g_n) \dots a^{\dag}(g_1)a(g_1)\dots a(g_n) \rho]}.
\end{equation}
First, note that due to the cyclic property of the trace
\begin{equation}
\tr[\Lambda_{\xi}(x)\rho_{-}] = \frac{\tr [a^{\dag}(g_n) \dots a^{\dag}(g_1)\Lambda_{\xi}(x)a(g_1)\dots a(g_n) \rho]}{{\rm tr}[a^{\dag}(g_n) \dots a^{\dag}(g_1)a(g_1)\dots a(g_n) \rho]}.
\end{equation}
By virtue of (\ref{eq:lambdaHeisSol}), we derive the following crucial identity in Appendix \ref{app:A}:
\begin{equation}\begin{split}\label{eq:important}
&a^{\dag}(g_n) \dots a^{\dag}(g_1) \Lambda_{\xi}(x) a(g_1)\dots a(g_n)\\
&=\Lambda_{\xi} [a^{\dag}(e^{\xi D}g_n) \dots a^{\dag}(e^{\xi D}g_1) x\,a(e^{\xi D}g_1)\dots a(e^{\xi D}g_n)],
\end{split}
\end{equation}
which holds for all $x\in {\cal A}$. By applying the identity (\ref{eq:identity}), we then obtain
\begin{align}
 \tr&[x\Lambda^{\star}_{\xi}(\rho_{-})]\\ &= \tr[\Lambda_{\xi}(x)\rho_{-}]\nonumber\\ &= \frac{\tr \{\Lambda_{\xi} [a^{\dag}(e^{\xi D}g_n) \dots a^{\dag}(e^{\xi D}g_1) x\,a(e^{\xi D}g_1)\dots a(e^{\xi D}g_n)] \rho\}}{{\rm tr}[a^{\dag}(g_n) \dots a^{\dag}(g_1)a(g_1)\dots a(g_n) \rho]}\nonumber\\
\label{eq:thisistrue} &= \frac{\tr \{ x\,a(e^{\xi D}g_1)\dots a(e^{\xi D}g_n) \Lambda^{\star}_{\xi} (\rho)a^{\dag}(e^{\xi D}g_n) \dots a^{\dag}(e^{\xi D}g_1)\}}{{\rm tr}[a^{\dag}(g_n) \dots a^{\dag}(g_1)a(g_1)\dots a(g_n) \rho]}\nonumber
\end{align}
Because these equalities hold for every observable $x \in {\cal A}$, we find that the action of the loss channel on the state $\rho_-$ is given by
\begin{equation}\label{eq:keyresult1}
\Lambda^{\star}_{\xi}(\rho_{-}) = \frac{a(e^{\xi D}g_1)\dots a(e^{\xi D}g_n) \Lambda^{\star}_{\xi} (\rho)a^{\dag}(e^{\xi D}g_n) \dots a^{\dag}(e^{\xi D}g_1)}{{\rm tr}[a^{\dag}(g_n) \dots a^{\dag}(g_1)a(g_1)\dots a(g_n) \rho]}.
\end{equation}
As such, we can relate the action of the loss channel on the photon-subtracted state $\rho_{-}$ to the action of the loss channel on the initial state $\rho$ from which the photons were subtracted. Nevertheless, we must also transform the creation and annihilation operators, such that losses and photon subtraction do not simply commute. \\

Even though this procedure shows how the loss channel acts on the photon subtracted state, it hardly provides any insight when presented in the form (\ref{eq:keyresult1}). In particular, the appearance of the operators $a^{\dag}(e^{\xi D}g_j)$ and $a(e^{\xi D}g_j)$ can incite confusion (see Appendix \ref{app:creation} for details). Hence, we will now recast (\ref{eq:keyresult1}) in a more insightful expression.

First of all, we can define the new vectors 
\begin{equation}\tilde{g}_j \equiv \frac{e^{\xi D}}{\norm{e^{\xi D} g_j}} g_j,\label{eq:gTilde}\end{equation} 
such that we can write that 
\begin{align}
&a^{\dag}(e^{\xi D}g_j) = \norm{e^{\xi D} g_j} a^{\dag}(\tilde{g}_j),\\
&a(e^{\xi D}g_j) = \norm{e^{\xi D} g_j} a(\tilde{g}_j).
\end{align}
When we insert this new notation in (\ref{eq:keyresult1}), {\em we find the following expression which is one of the main results of this article}:
\begin{equation}\label{eq:lossFinal}
\Lambda^{\star}_{\xi}(\rho_-) = \frac{a(\tilde{g}_1)\dots a(\tilde{g}_n) \Lambda^{\star}_{\xi} (\rho) a^{\dag}(\tilde{g}_n) \dots a^{\dag}(\tilde{g}_1) }{{\rm tr}[a^{\dag}(\tilde{g}_n) \dots a^{\dag}(\tilde{g}_1)a(\tilde{g}_1)\dots a(\tilde{g}_n) \Lambda_{\xi}^{\star} (\rho) ]},
\end{equation}
where we used that 
\begin{align}
&\left(\prod_{j=1}^n\norm{e^{\xi D} g_j}^2\right){\rm tr}[a^{\dag}(\tilde{g}_n) \dots a^{\dag}(\tilde{g}_1)a(\tilde{g}_1)\dots a(\tilde{g}_n) \Lambda_{\xi}^{\star} (\rho) ]\\
&=\left(\prod_{j=1}^n\norm{e^{\xi D} g_j}^2\right){\rm tr}\{\Lambda_{\xi}[a^{\dag}(\tilde{g}_n) \dots a^{\dag}(\tilde{g}_1)a(\tilde{g}_1)\dots a(\tilde{g}_n)] \rho \}\nonumber\\
&={\rm tr}[a^{\dag}(g_n) \dots a^{\dag}(g_1)a(g_1)\dots a(g_n) \rho].\nonumber
\end{align}
This implies that subtracting $n$ photons from a state $\rho$ in modes $g_1, \dots, g_n$, and subsequently having losses characterised by $\Lambda^{\star}_{\xi}$ is equivalent to subtracting $n$ photons from the state $\Lambda^{\star}_{\xi} (\rho)$ in modes $\tilde{g}_1, \dots, \tilde{g}_n$. 
%On the one hand, losses ultimately destroy non-Gaussian effects due to $\Lambda_{\xi \rightarrow \infty}' (\rho) = \lvert0\rangle\langle0\rvert$. 
Mode-dependent losses can also change the mode structure of the subtracted photons. This can be of particular interest in terms of generating multimode photon subtracted states that are more robust against such losses. The detrimental effect of strong losses on non-Gaussian features can be understood from the fact that $\Lambda_{\xi \rightarrow \infty}^{\star} (\rho^-) = \lvert0\rangle\langle0\rvert$ in (\ref{eq:lossFinal}). 

In the case where the losses are the same for all modes --which is quite common in experiments-- the expression simplifies considerably. Here we have $\exp(\xi D) = \exp(\xi \gamma/2) \mathds{1}$, which is equivalent to the beamsplitter model (\ref{eq:beamsplitterManyMode}). One directly obtains that $\tilde{g}_j = g_j$ for every mode in the mode basis, such that the photon subtractions and the noise channel commute. This means that uniform losses from a photon-subtracted state are equivalent to subtracting photons from a state that has undergone the same losses.\\

\subsection{Conceptual approach}\label{sec:concept}

Section \ref{sec:AlgApproach} makes it mathematically evident that photon subtraction and loss commute under the condition that $\tilde{g}_j = g_j$. In this section, we strive to provide a conceptual explanation for these findings through the beamsplitter model of Section \ref{sec:BeamsplitterModel}. In a single-mode setup, the condition for commutation between loss and photon subtraction is always fulfilled, and, thus, it is instructive to start our conceptual treatment with this simple case.

\begin{figure}
\centering
\includegraphics[width=0.49\textwidth]{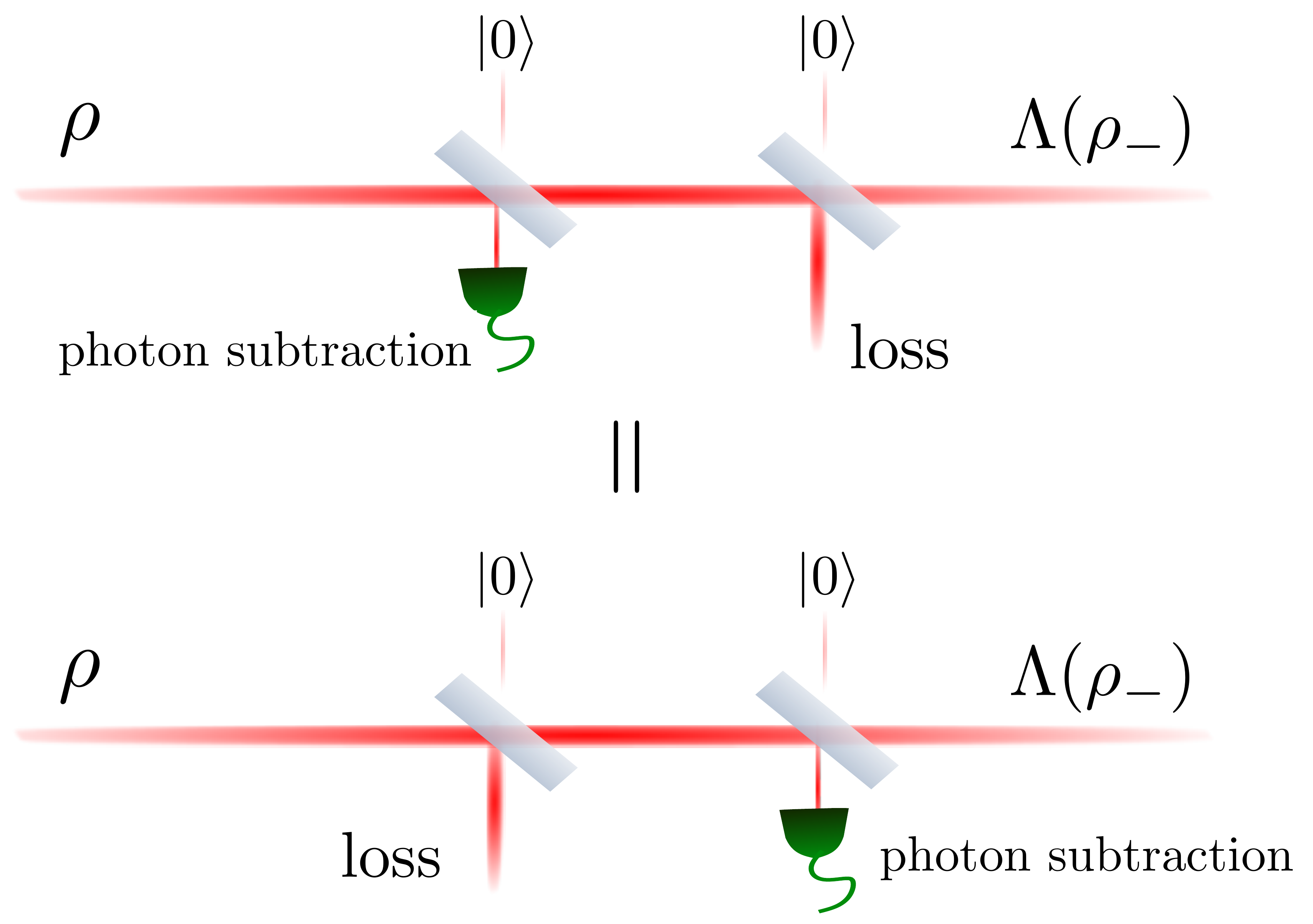}
\caption{Schematic representation of single-mode photon subtraction (as implemented by a highly transmitting beamsplitter) and a photodetector, and single-mode losses (represented by a beamsplitter). The top and bottom configuration are shown to be equivalent (see main text).\label{fig:LossesBS}}
\end{figure}

In the single-mode case, photon subtraction is generally implemented using a beamsplitter with a very low reflectivity and a photodetector, as shown in Fig.~\ref{fig:LossesBS}. The beamsplitter reflects a minor portion of the light to the photodetector, which then heralds a photon-subtracted state in the transmitted beam upon detection. This heralding procedure effectively implements the annihilation operator. As explained in detail in Section \ref{sec:BeamsplitterModel}, the losses can be modelled by a beamsplitter, which is also represented in Fig.~\ref{fig:LossesBS}.

To understand why the top and bottom panel of Fig.~\ref{fig:LossesBS} give rise to the same output state, we first focus on the bottom panel, where the photon is subtracted after the losses occur. Here, the state $\rho$ is first mixed with a certain amount of vacuum in the loss process, and subsequently, the photodetector heralds the photon-subtracted state. To conceptually understand the process, we can trace back to origin of the subtracted photon. Going back to the stage that implements the losses, we see that the photon can either originate from the initial state $\rho$, or from the other input port of the beamsplitter that inserts the vacuum component. However, the vacuum $\lvert 0 \rangle$ is an eigenstate of the number operator that contains exactly zero photons. Hence, the subtracted photon cannot possibly originate from the $\lvert 0 \rangle$ input in the loss-beamsplitter and must, therefore, stem from the state $\rho$. This means that it is equivalent to subtracting a photon prior to the losses. Note, furthermore, that photon-subtraction is a probabilistic operation, with a higher success probability before the losses than after. Because the final state is conditioned on a successful detection event (hence the need to renormalise the state after applying the annihilation operator), this difference in success probability plays no role in the final state. This conceptually explains Fig.~\ref{fig:LossesBS} in agreement to the algebraic derivation of Section \ref{sec:AlgApproach}\\

\begin{figure}
\centering
\includegraphics[width=0.49\textwidth]{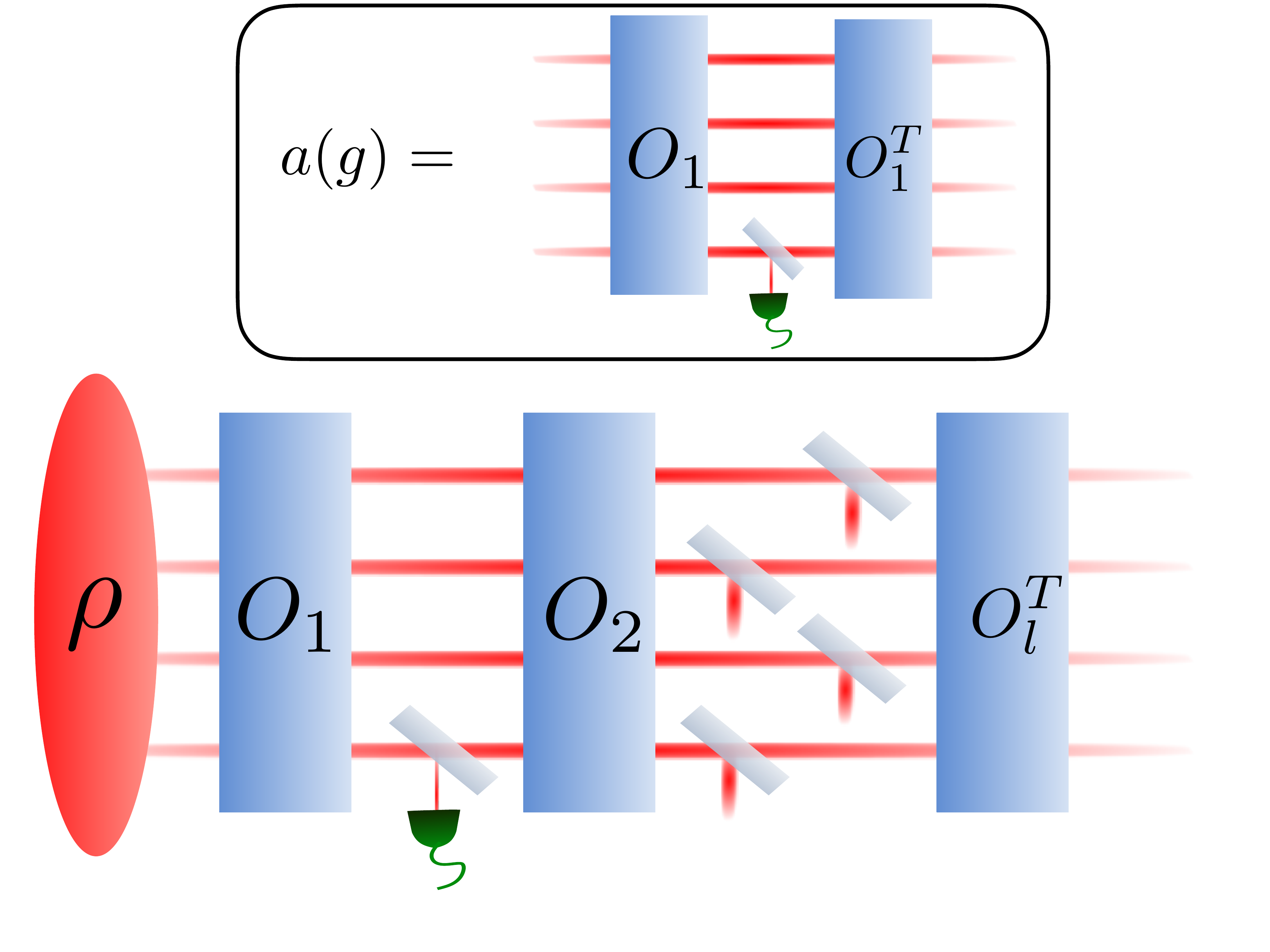}
\caption{Schematic representation of mode dependent losses (see also Fig.~\ref{fig:LossesIllust}) acting on a photon-subtracted state, where we have defined $O_2 = O_1^T O_l$. Inset shows the schematic representation of mode-selective photon subtraction. \label{fig:LossesMultimode}}
\end{figure}

In the multimode scenario, a significant layer of complexity is added when the losses and and the photon subtraction can all act in different modes bases. In Section \ref{sec:BeamsplitterModel}, we explained how the beamsplitter point of view can be adopted when additional basis changes are included as shown in Fig.~\ref{fig:LossesIllust}. A similar type of logic can be applied to photon subtraction in a general mode $g \in {\cal N}(\mathbb{R}^{2m})$ as shown in the inset of Fig.~\ref{fig:LossesMultimode}: the annihilation operator $a(g)$ can then be represented as a basis change $O_1$ which translates the mode basis in which the multimode state $\rho$ is expressed to a mode basis that contains the subtraction mode $g$ as a basis vector. The second mode basis change $O_1^T$ is required to revert the photon-subtracted state to the initial mode basis. Note that $O_1$ is certainly not unique, because there are many mode basis that contain $g$.

Mode-dependent losses and mode-dependent photon subtraction are combined in Fig.~\ref{fig:LossesMultimode}. Of particular interest is the appearance of the mode basis change $O_2 = O_1^T O_l$, which generally is not a trivial transformation. This transformation makes it impossible to repeat the argument that was used in Fig.~\ref{fig:LossesBS} for commuting losses and photon subtraction. However, when the losses and photon subtraction act in the same mode basis, we find that $O_2 = \mathds{1}$. In that scenario, the logic of Fig.~\ref{fig:LossesBS} holds, and the losses commute with the photon subtraction. \\ 

These results are conceptually interesting, but for a more quantitative understanding we must understand the action of the loss channel on $\rho$, and we must be able to calculate the multimode photon-subtracted state. Both of these aspects are not for granted, and therefore we focus on the specific case of single-photon subtraction from a Gaussian state in the following section.

%Because the creation and annihilation operators are generators of the algebra\footnote{Note, more accurately, they generate any operator in the Fock representation of the algebra.} of observables ${\cal A}$, we can approximate any observable $x \in {\cal A}$ by a polynomial of creation and annihilation operators. Through application of the canonical commutation relations, we can then cast all terms in this polynomial in normal order, to obtain the series expansion
%\begin{equation}\label{eq:xExpansion}
%x = \sum_{n_1, \dots, n_m = 0}^{\infty}\sum_{n'_1, \dots, n'_m = 0}^{\infty}  X^{n_1\dots n_m}_{n'_1 \dots n'_m}\, (a^{\dag}_1)^{n_1}\dots (a^{\dag}_m)^{n_m}(a_m)^{n'_1}\dots (a_1)^{n'_m},
%\end{equation}
%where $a^{\dag}_1, \dots, a^{\dag}_m$ and $a_1, \dots, a_m$ are creation and annihilation operators, respectively, for a randomly chosen mode basis, and $X^{n_1\dots n_m}_{n'_1 \dots n'_m} \in \mathbb{C}$ are the coefficients of the polynomial which represents $x\in {\cal A}$.
%Hence, the result in (\ref{eq:lambdaHeisSol}) can be used to describe the full action of the channel on an arbitrary observable $x$. 

\section{Single-photon subtracted Gaussian states}\label{sec:result1photon}

\subsection{General results}

In this section, we focus on the effect of losses on a Gaussian state $\rho_G$ with a single photon subtracted from it. As we mentioned in Section \ref{sec:lossopen}, Gaussian states are fully characterised by the expectation values and pair-correlations of quadrature measurements. Here, we will assume that the state is not displaced, such that $\tr[Q(f)\rho_G] = 0$ for all $f \in {\cal N}(\mathbb{R}^{2m})$, with the quadrature operator $Q(f)$ as defined in (\ref{eq:Qgen}). This implies that the state is fully characterised by its covariance matrix $V$ that captures all the pair-correlations.

Photon subtraction will render the state non-Gaussian, such that a covariance matrix alone is no longer sufficient to describe the photon-subtracted state. In this context, it is often convenient to use the Wigner function as general state representation \cite{PhysRev.40.749,PhysRev.177.1882}. In the case of a non-displaced Gaussian, we can write the Wigner function as
\begin{equation}\label{eq:wigGauss}
W_G(\beta) =  \frac{e^{-\frac{1}{2}(\beta, {V}^{-1} \beta)}}{(2\pi)^n \sqrt{\det V}},
\end{equation}
where $\beta \in \mathbb{R}^{2m}$ is an arbitrary point in the optical phase space. In the case of Gaussian states, the Wigner function can be interpreted as a joint probability distribution for the outcomes of quadrature measurements. In general, this interpretation does not hold since the Wigner function can take negative values. Exactly this behaviour can be induced through photon subtraction.

When a single photon is subtracted from such a Gaussian state $\rho_G$, we write the new state as
\begin{equation}
\rho_- = \frac{a(g) \rho_G a^{\dag}(g)}{\tr [a^{\dag}(g)a(g) \rho_G]}.
\end{equation}
We showed in \cite{walschaers_statistical_2017,walschaers_entanglement_2017} that the Wigner function of $\rho_-$ can be obtained analytically, and is given by
\begin{equation}\label{eq:wig}
W_-(\beta)=\frac{1}{2}\big[ (\beta, V^{-1}A_gV^{-1} \beta) -{\rm tr}\{V^{-1}A_g\}+2 \big]W_G(\beta),
\end{equation}
where, $W_G$ is given by (\ref{eq:wigGauss}), and 
\begin{equation}\label{eq:Ag}
A_g = 2\frac{(V-\mathds{1})(P_g+P_{Jg})(V-\mathds{1})}{{\rm tr}\{(V-\mathds{1})(P_g + P_{Jg})\}}.
\end{equation}
The matrix $A_g$ is a rank-two matrix, which is narrowly related to the quadrature correlations in the photon-subtracted state.

With (\ref{eq:lossFinal}), we find that for our loss model, the photon subtracted state is changed according to
\begin{equation}\label{eq:lossFinalOneSub}
\Lambda_{\xi}^{\star}(\rho_-) = \frac{a(\tilde{g}) \Lambda_{\xi}^{\star}(\rho_G) a^{\dag}(\tilde{g})}{\tr [a^{\dag}(\tilde{g})a(\tilde{g}) \Lambda_{\xi}^{\star}(\rho_G)]},
\end{equation}
where $\tilde{g}=\exp(\xi D) g / \norm{\exp(\xi D) g}$. The Wigner function of this state can then directly be obtained as
\begin{equation}\label{eq:wigloss}
W^{\xi}_-(\beta)=\frac{1}{2}\big[ (\beta, V_{\xi}^{-1}A^{\xi}_{\tilde{g}} V_{\xi}^{-1} \beta) -{\rm tr}\{V_{\xi}^{-1}A^{\xi}_{\tilde{g}}\}+2 \big]W^{\xi}_G(\beta),
\end{equation}
where $V_{\xi}$ is given by (\ref{eq:Vxi}), and $W^{\xi}_G$ is the Wigner function of the Gaussian state upon which the loss channel has acted:
\begin{equation}\label{eq:wigGaussLoss}
W^{\xi}_G(\beta) =  \frac{e^{-\frac{1}{2}(\beta, {V}_{\xi}^{-1} \beta)}}{(2\pi)^n \sqrt{\det V_{\xi}}}.
\end{equation}
Furthermore, the non-Gaussian features are induced by the matrix
\begin{equation}
A^{\xi}_{\tilde{g}} = 2\frac{(V_{\xi}-\mathds{1})(P_{\tilde{g}}+P_{J\tilde{g}})(V_{\xi}-\mathds{1})}{{\rm tr}\{(V_{\xi}-\mathds{1})(P_{\tilde{g}} + P_{J\tilde{g}})\}}.
\end{equation}

It can be more convenient to recast the expression for $A^{\xi}_{\tilde{g}}$. When we combine $(V_{\xi} - \mathds{1}) = e^{-\xi D} (V-\mathds{1}) e^{-\xi D}$ with $\tilde{g}=\exp(\xi D) g / \norm{\exp(\xi D) g}$, we find that we can rewrite 
\begin{equation}\label{eq:Agfinal}
A^{\xi}_{\tilde{g}} = e^{-\xi D}A_ge^{-\xi D},
\end{equation}
which transforms the Wigner function of the photon-subtracted state in a lossy channel to the more insightful expression
\begin{align}
&W^{\xi}_-(\beta)\\
&=\frac{1}{2}\big[ (\beta, V^{-1}_{\xi}e^{-\xi D}A_{g}e^{-\xi D}V^{-1}_{\xi} \beta) -{\rm tr}\{V^{-1}_{\xi}e^{-\xi D}A_ge^{-\xi D}\}+2 \big]\nonumber\\
&\qquad\times W^{\xi}_{G}(\beta).\nonumber
\end{align}
Because all the non-Gaussian features are induced by $\exp(-\xi D)A_{g}\exp(-\xi D),$ we clearly see that these features vanish for increasing $\xi$. In particular, we find that for the limit $\xi \rightarrow \infty$, the state converges to the vacuum (assuming that there are no decoherence free subspaces, which is equivalent to demanding that $D$ is invertible).

We can use the same reasoning as in \cite{walschaers_statistical_2017,walschaers_entanglement_2017} to obtain a more strict condition for the existence of negative values of the Wigner function. Due to the positivity of the matrix $V^{-1}_{\xi}e^{-\xi D}A_{g}e^{-\xi D}V^{-1}_{\xi}$, we find the necessary and sufficient condition
\begin{equation}
{\rm tr}\{V^{-1}_{\xi}e^{-\xi D}A_ge^{-\xi D}\} > 2.
\end{equation}
This condition can be rewritten as
\begin{equation}\label{eq:negcondnew}\begin{split}
&(g,e^{\xi D}V^{-1}_{\xi}e^{\xi D} g) +(Jg,e^{\xi D}V^{-1}_{\xi}e^{\xi D}  Jg) \\
&> (g, e^{2\xi D} g) +(J g, e^{2\xi D} J g).
\end{split}
\end{equation}
This clearly shows that there is a loss threshold for the negativity of the Wigner function, and that it is reached very quickly.

\subsection{Examples}
\paragraph{First example} We illustrate the usefulness of our loss model by means of several examples. First, we study the impact of losses on single-mode photon subtracted states, which leads us to a scenario that is similar to \cite{biswas_nonclassicality_2007}. Typically, single mode quantum optics experiments aim at optimising the purity of the generated quantum states, such that a squeezed vacuum with losses accurately captures the experimental reality. However, in general there may also be thermal noise present in the mode (e.g., when the mode is actually entangled to other modes). In general, we can always tune the phase reference for the phase and amplitude quadratures such that the covariance matrix of a Gaussian state is given by
%However, in multimode setups it is quite common to have large sets of entangled modes. When we then measure just a single mode from this larger set, the entanglement will effectively induce thermal noise in the reduced single mode state. When this multimode state is a squeezed vacuum, we find that the reduced state for a single mode is a Gaussian state with a covariance matrix of the form
%However, a general Gaussian single mode state can be described by a thermal state which is squeezed and displaced. 
\begin{equation}\label{eq:Vsinglemode}
V = \begin{pmatrix} n s & 0 \\ 0 & n s^{-1} \end{pmatrix}.
\end{equation} 
where $s$ describes squeezing and $n$ the thermal noise. Here, we consider photon subtraction from such a state, and study the effect of a subsequent loss channel, governed by parameter $\xi$. 

\begin{figure}
\centering
\includegraphics[width=0.49\textwidth]{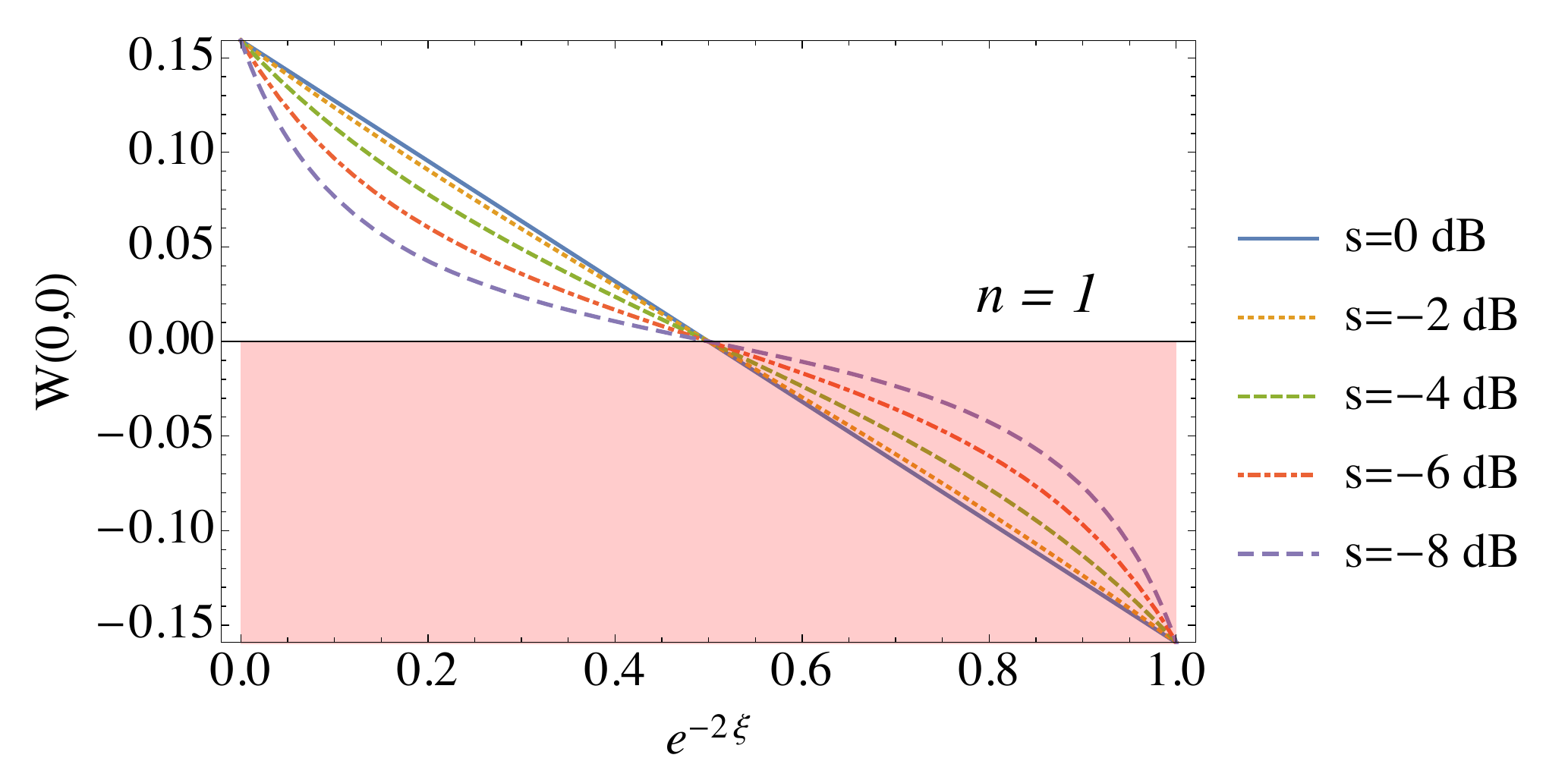}
\includegraphics[width=0.45\textwidth]{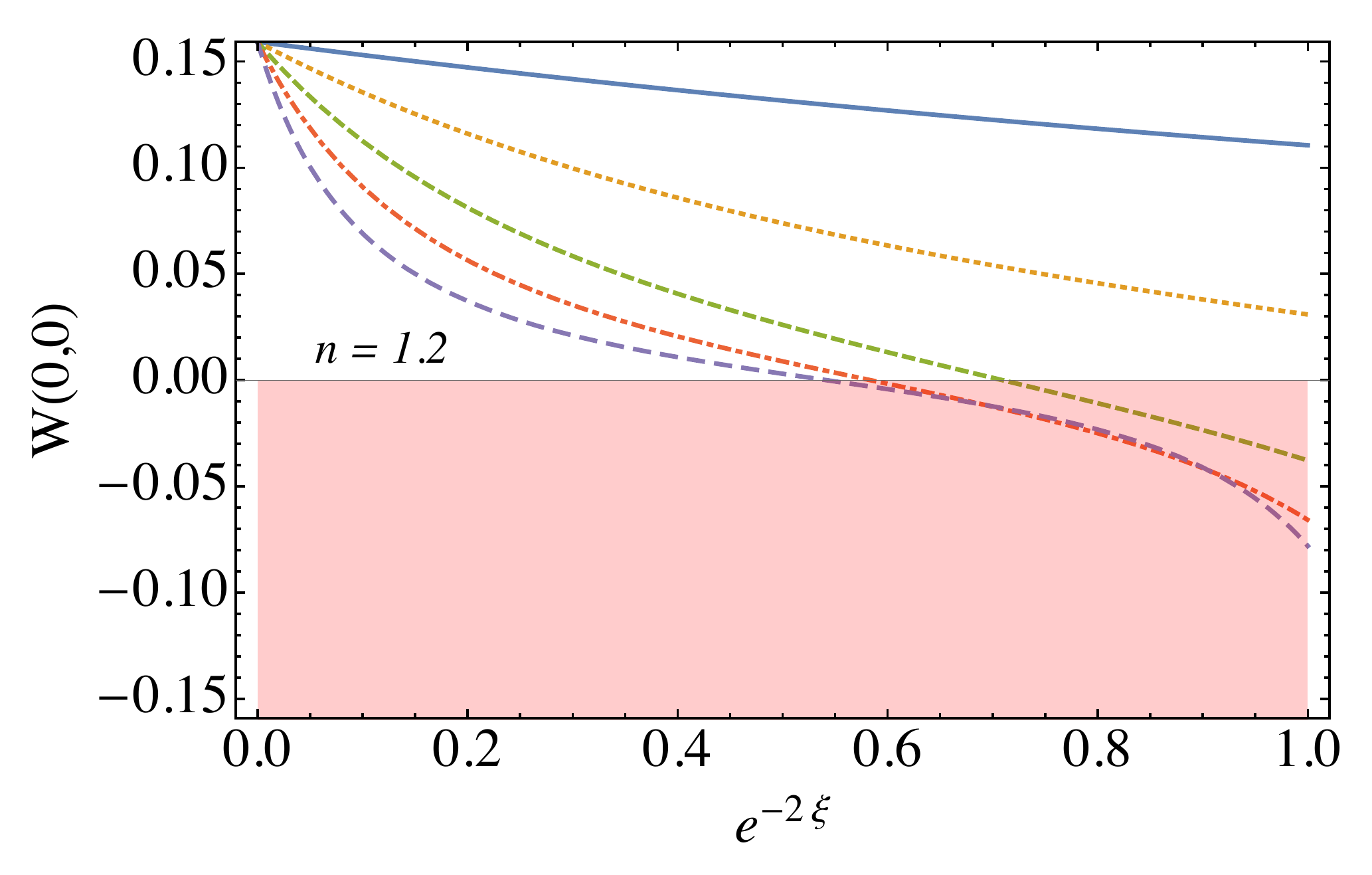}
\caption{Value of a single-mode Wigner function in the origin of phase space as a function of the strength of the losses, probed by $\exp(-2\xi)$ in (\ref{eq:wigloss}). Different values of squeezing are explored (see legend) for, both, photon subtraction from a pure state with $n=1$ in (\ref{eq:Vsinglemode}), and a noisy state with $n=1.2$. The highlighted area indicates a negative value for the Wigner function. \label{fig:neg1mode}}
\end{figure}

Note that in the single mode regime, the loss model (\ref{eq:wigloss}) is considerably simplified: we find that $D= \mathds{1}$, such that 
\begin{align}
&V_{\xi} = \mathds{1} + e^{-2\xi} (V - \mathds{1}),\label{eq:singleModeCovLoss}\\
&A^{\xi}_{\tilde{g}} = e^{-2\xi} A_{g}. \label{eq:singleModeNonGauss}
\end{align}
Losses generally have a detrimental effect on the negativity of the Wigner function, which can directly be traced back to (\ref{eq:singleModeNonGauss}). The effect of losses on the state's Wigner function (\ref{eq:wigloss}) as evaluated in the origin of phase space is shown in Fig.~\ref{fig:neg1mode}. The top panel shows the effect of losses on a photon-subtracted squeezed vacuum, i.e.~$n=1$. We find that, independent of the amount of squeezing, the Wigner function is positive when $e^{- 2\xi} \leqslant  1/2$. On the other hand, for all $\xi$ with $e^{- 2\xi} > 1/2$, we find that the Wigner function does reach negative values. In other words, the the negativity of the Wigner function vanishes at exactly $50\%$ loss. These observations are in perfect agreement with existing literature \cite{biswas_nonclassicality_2007}. Furthermore, higher squeezing values are seen to be more sensitive to noise, as the minimal value of the Wigner function $W(0,0)$ increases faster with the losses.

When set $n>1$ in (\ref{eq:Vsinglemode}), as in the bottom panel in Fig.~\ref{fig:neg1mode}, we still find that the Wigner function is positive for $e^{- 2\xi} \leqslant  1/2$. The origin of this effect can be seen from (\ref{eq:singleModeCovLoss}): whenever $e^{- 2\xi} \leqslant  1/2$, we find that more than half of the state is made up of vacuum. However, due to the additional thermal noise, it is no longer true that smaller amounts of noise automatically lead to non-positive Wigner functions. In particular, we now find that for a fixed value of thermal noise $n$ the Wigner functions of photon-subtracted states with more squeezing are more robust to losses. Hence, this is in line with the findings in \cite{walschaers_statistical_2017}, where we showed the interplay between thermal noise and squeezing. A thermal state needs to be sufficiently squeezed for the photon-subtraction to induce negativity in the Wigner function. Moreover, in the single mode case $g=\tilde g$ in (\ref{eq:lossFinalOneSub}), such that the losses and the photon subtraction commute, and the behaviour of the lossy photon-subtracted state can be understood from the structure (\ref{eq:singleModeCovLoss}).

\paragraph{Second example} The situation becomes considerably more interesting when multimode states and mode dependent-losses are considered. We recently showed that photon subtraction from an entangled state can affect multiple modes, which can be nicely illustrated by studying photon subtraction from CV graph states \cite{Walschaers-graphs-2018}. In this second example, we will investigate how losses affect these states, in particular when these losses are mode-dependent.

First, let us present a brief introduction to CV graph states. To construct a graph state, we follow the recipe of \cite{gu_quantum_2009}, where these states are constructed by applying a network of $C_Z$ gates to a multimode squeezed vacuum with infinite squeezing. In more realistic setups \cite{PhysRevA.76.032321,cai-2017,roslund_wavelength-multiplexed_2014,PhysRevLett.112.120505}, one has to make due with a finite amount of squeezing. Therefore, we follow the formalism of \cite{PhysRevA.83.042335} to describe graph states with finite squeezing. In particular, we start from an initial multimode squeezed vacuum with a covariance matrix
\begin{equation}
V_0 = {\rm diag}(s_1, \dots, s_m, s_1^{-1}, \dots, s_m^{-1}),
\end{equation}
upon which we act with a series of $C_Z$ gates, defined by the unitary operation $\hat{C}_Z = \exp(i \hat{x}_i \hat{x}_j)$ when it is applied on modes with labels $i$ and $j$. Because both the initial state and the $C_Z$ gate are Gaussian, we can describe the graph state fully on the level of its covariance matrix $V$, which is constructed as
\begin{equation}\label{eq:clusterV}
V = G^tV_0G, \text{ with } G= \begin{pmatrix} \mathds{1}& {\cal A}\\ 0  & \mathds{1}\end{pmatrix},
\end{equation}
where $G$ is the symplectic transformation that describes the application of the $C_Z$ gates, as prescribed by the graph's connectivity matrix ${\cal A}$: when the component ${\cal A}_{ij} = 1$, the modes $i$ and $j$ are entangled by a $C_Z$ gate.

Photon-subtraction from such states can, again, be described through eq.~(\ref{eq:wig}), the consequences of which are analysed in \cite{Walschaers-graphs-2018} and experimentally realised in \cite{Ra2019}. In particular, we highlighted that the spread of non-Gaussian features can be understood via the single-mode Wigner functions for each one of the vertices. We follow the same strategy when we study the impact of losses on these photon-subtracted graph states. As these examples mainly serve as an illustration of the most relevant phenomena, we can safely limit ourselves to small systems: in the present case square graphs. The photon is subtracted locally, in one of the vertices of the graph, and, as shown in in \cite{Walschaers-graphs-2018,Ra2019}, this affects the square in its entirety. 

\begin{figure*}
\centering
\includegraphics[width=0.8\textwidth]{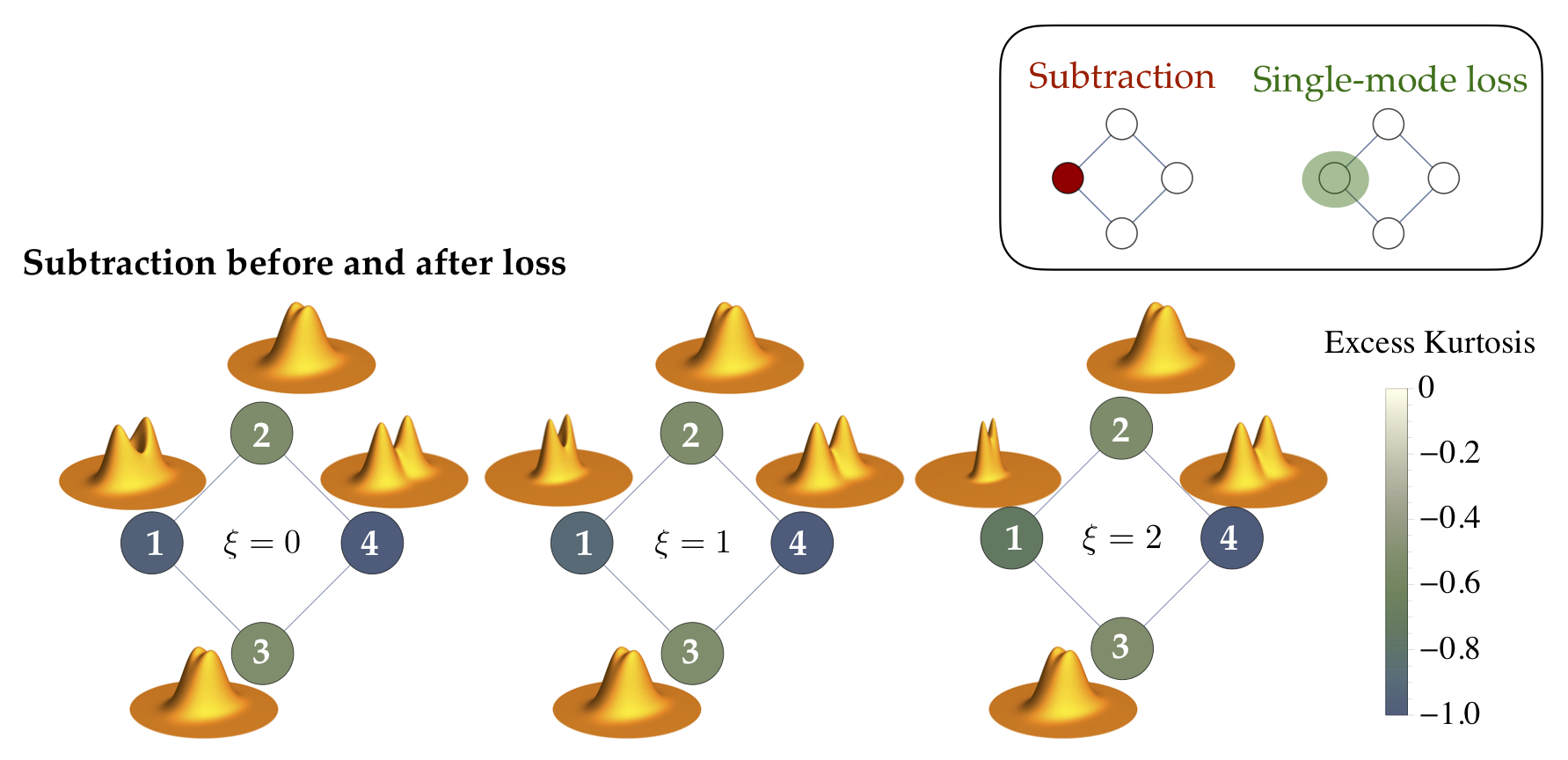}
\caption{Graph state (\ref{eq:clusterV}) of four vertices (modes), with a photon subtracted in the mode associated with the red vertex (inset), probed for different degrees $\xi$ of losses. The single-mode reduced Wigner function is shown for each vertex $k$. The minimal excess kurtosis (\ref{eq:kurt}) in each vertex is represented by the color code. Losses (\ref{eq:SpectTheorem}) act on a single vertex of the graph, as indicated in the inset (see main text for details). All modes in the initial squeezed vacuum $V_0$ are equally squeezed (i.e.~$s_1= \dots =s_m$) at $10{\rm dB}$.\label{fig:NodeLossGraph}}
\end{figure*} 

\begin{figure*}
\centering
\includegraphics[width=0.8\textwidth]{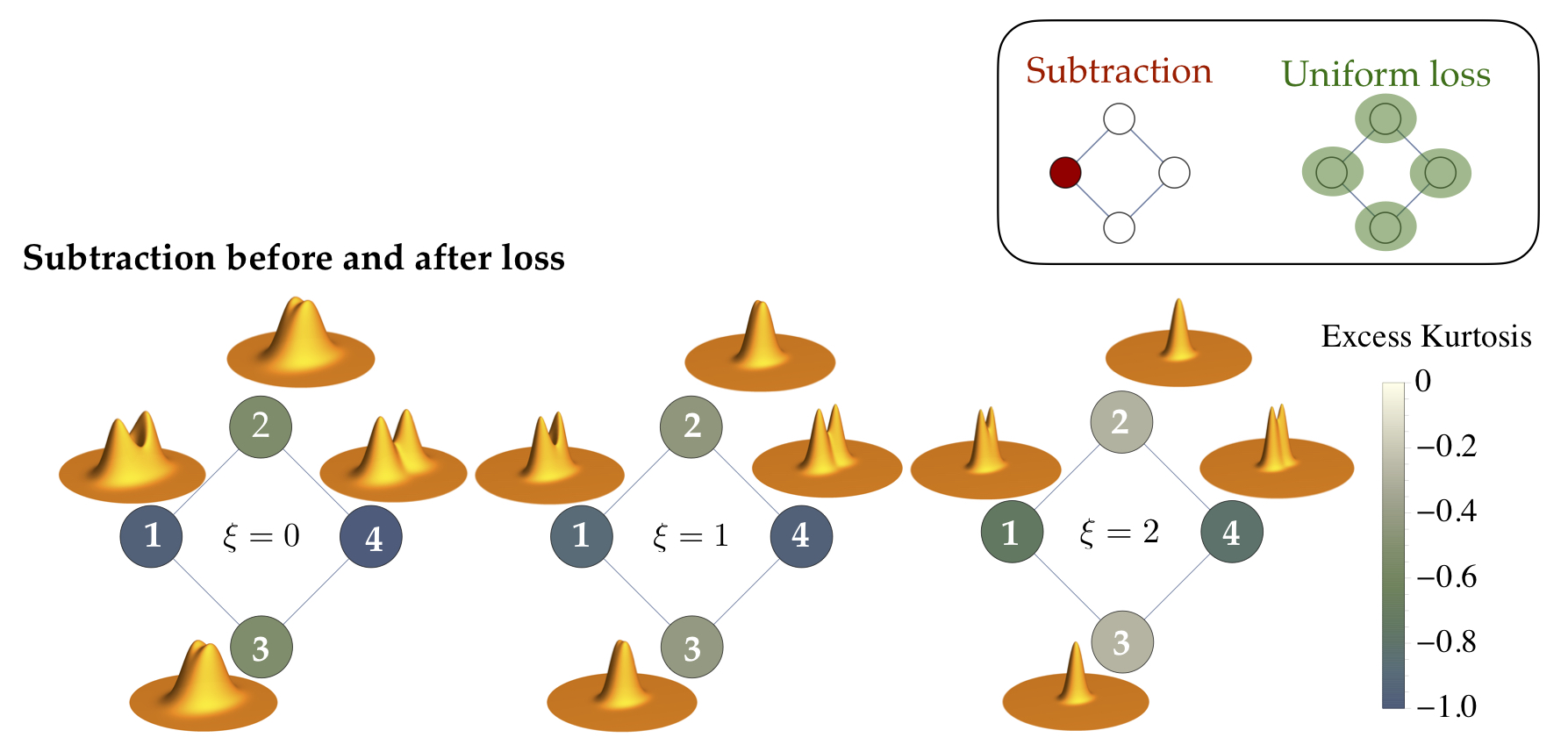}
\caption{Graph state (\ref{eq:clusterV}) of four vertices (modes), with a photon subtracted in the mode associated with the red vertex (inset), probed for different degrees $\xi$ of losses. The single-mode reduced Wigner function is shown for each vertex $k$. The minimal excess kurtosis (\ref{eq:kurt}) in each vertex is represented by the color code. Losses (\ref{eq:SpectTheorem}) act in a uniform way on all vertices, as indicated in the inset (see main text for details). All modes in the initial squeezed vacuum $V_0$ are equally squeezed (i.e.~$s_1= \dots =s_m$) at $10{\rm dB}$.\label{fig:UniformLossGraph}}
\end{figure*}

\begin{figure*}
\centering
\includegraphics[width=0.8\textwidth]{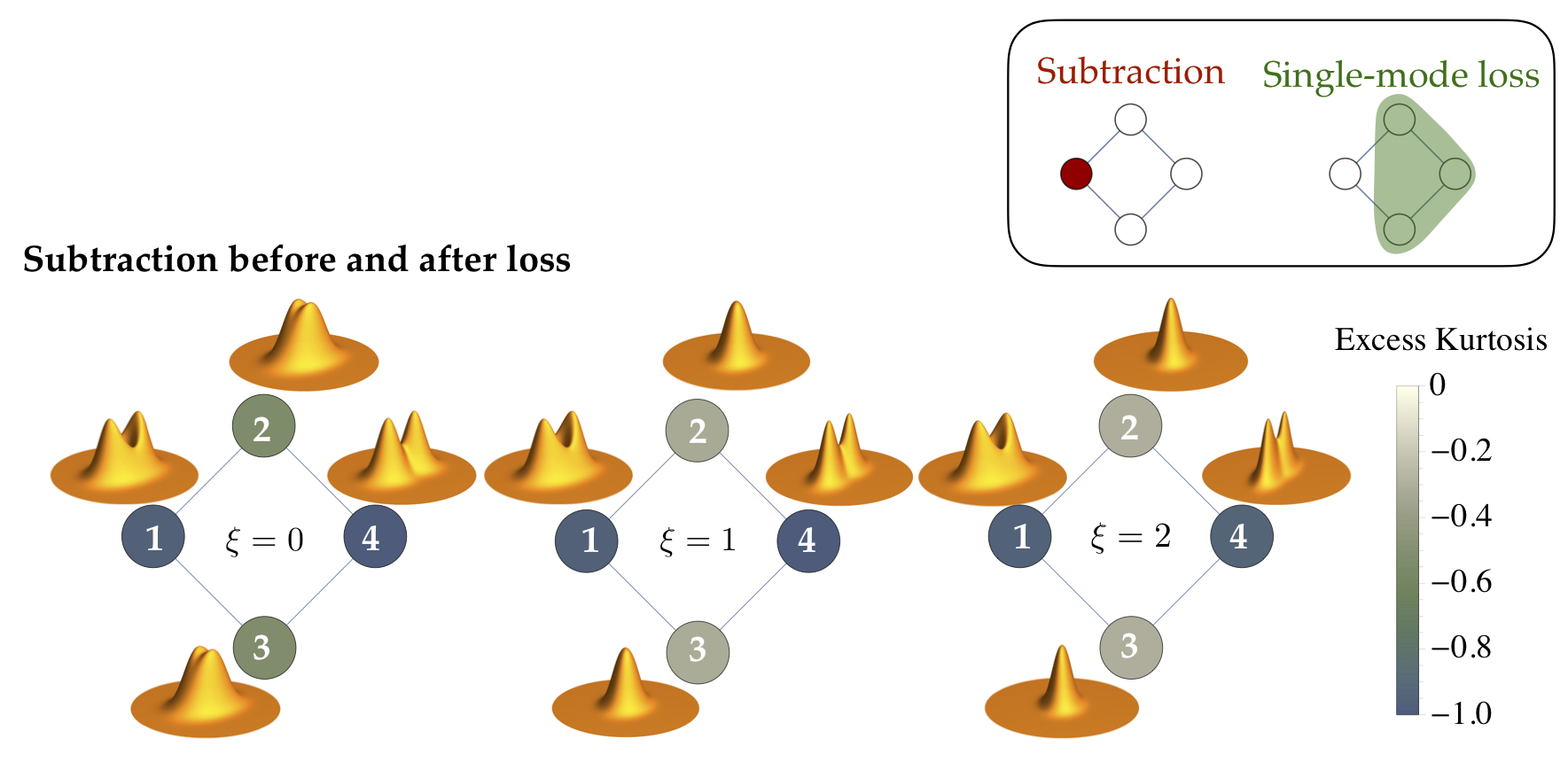}
\caption{Graph state (\ref{eq:clusterV}) of four vertices (modes), with a photon subtracted in the mode associated with the red vertex (inset), probed for different degrees $\xi$ of losses. The single-mode reduced Wigner function is shown for each vertex $k$. The minimal excess kurtosis (\ref{eq:kurt}) in each vertex is represented by the color code. Losses (\ref{eq:SpectTheorem}) act on a single mode that is a superposition of several vertices, as indicated in the inset (see main text for details). All modes in the initial squeezed vacuum $V_0$ are equally squeezed (i.e.~$s_1= \dots =s_m$) at $10{\rm dB}$.\label{fig:Local-Comp}}
\end{figure*}

\begin{figure*}
\centering
\includegraphics[width=0.8\textwidth]{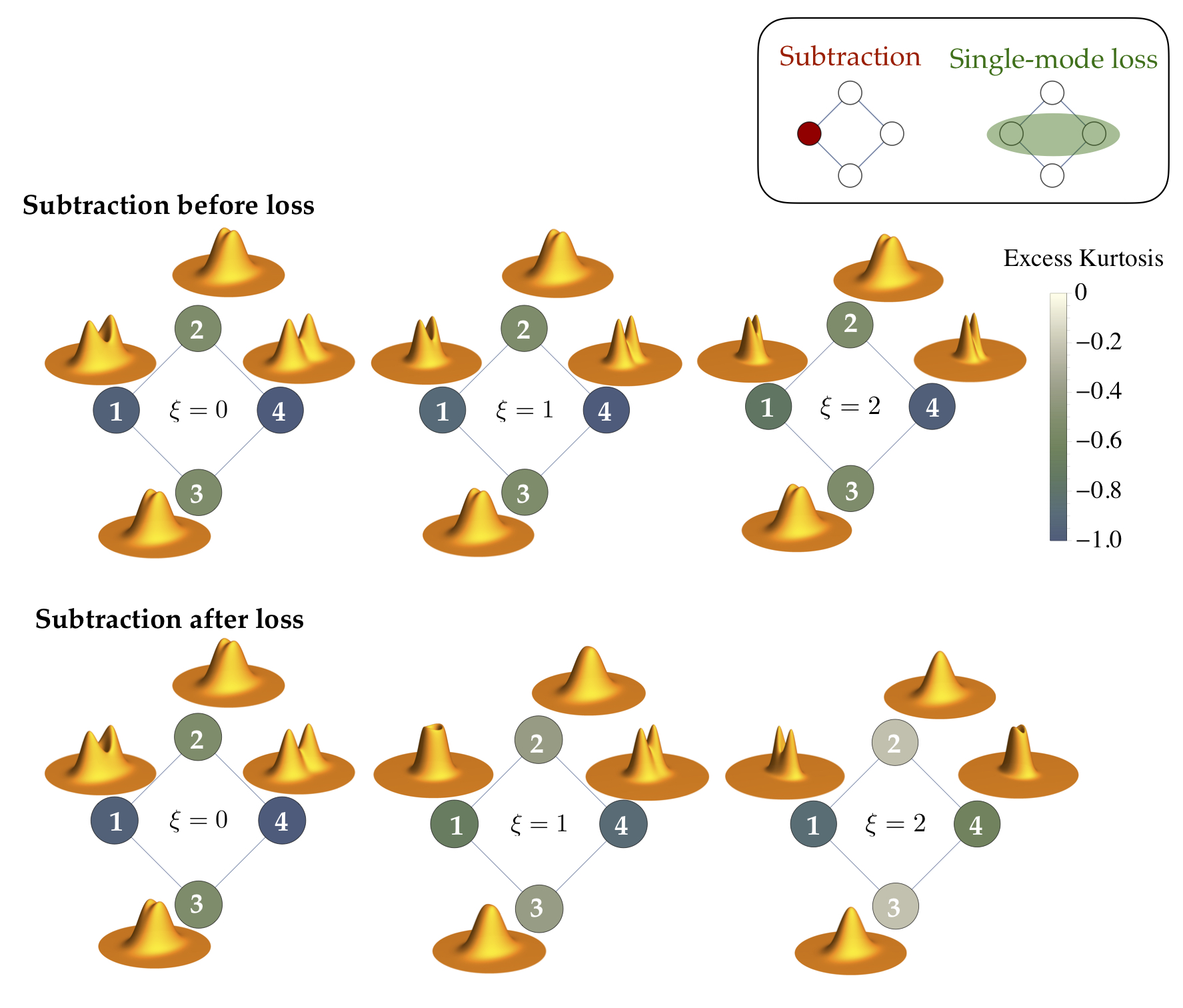}
\caption{Graph state (\ref{eq:clusterV}) of four vertices (modes), with a photon subtracted in the mode associated with the red vertex (inset), probed for different degrees $\xi$ of losses. The single-mode reduced Wigner function is shown for each vertex $k$. The minimal excess kurtosis (\ref{eq:kurt}) in each vertex is represented by the color code. Losses (\ref{eq:SpectTheorem}) act on a single mode that is a superposition of several vertices, as indicated in the inset (see main text for details). Losses and photon subtraction do not commute, hence we show both scenarios: subtraction prior to losses (top), and losses prior subtraction (bottom). All modes in the initial squeezed vacuum $V_0$ are equally squeezed (i.e.~$s_1= \dots =s_m$) at $10{\rm dB}$.\label{fig:ModeLossGraph}}
\end{figure*}

In Figs.~\ref{fig:NodeLossGraph}-\ref{fig:ModeLossGraph}, we subtract the photon in the leftmost vertex. The case where $\xi = 0$ corresponds to the scenario in the absence of losses, where we clearly see that non-Gaussian features are present in all single-mode Wigner functions. However, it is remarkable that these features are most pronounced in the vertex where the photon is subtracted, and in its next-to-nearest neighbour. The colours of the different nodes of the graph indicate the {\em minimal excess kurtosis} $\kappa_{\min}(f)$ in the mode, as given by
\begin{equation}\label{eq:kurt}
\kappa_{\min}(f) = \min_{\theta \in [0,2\pi]} \frac{\tr [Q(f_{\theta})^4 \Lambda^{\star}_{\xi}(\rho_-)]}{\tr [Q(f_{\theta})^2 \Lambda^{\star}_{\xi}(\rho_-)]^2} - 3
\end{equation}
where $f_{\theta} = \cos (\theta) f + \sin(\theta) Jf$. The minimal excess kurtosis serves as a measure for non-Gaussianity: for Gaussian states it is exactly zero. Values $\kappa_{\min}(f) > 0$ indicate that the tail of the distribution is heavier than a Gaussian distribution, whereas $\kappa_{\min}(f) < 0$ implies a sub-Gaussian tail \cite{Westfall:2014aa}. We showed \cite{walschaers_statistical_2017} that photon-subtracted Gaussian states have a negative excess kurtosis for at least one quadrature $Q(f)$, which is why, to assess the most non-Gaussian feature, 
%we choose to minimize in (\ref{eq:kurt}). Even a single mode has infinitely many marginals that can be measured through varying the phase. Hence, 
%to assess the most non-Gaussian feature, 
%consistent with photon subtraction, in our single mode Wigner function, 
we probe the phase that leads to the smallest, i.e.~most negative, value for the kurtosis. Lower values for the kurtosis are indicated by darker colours in the nodes of the graphs of Figs.~\ref{fig:NodeLossGraph}-\ref{fig:ModeLossGraph}. %which confirms that the mode of subtraction, and the one two vertices away from it have the most pronounced non-Gaussian features.
The middle and rightmost graphs in these figures show the impact of a loss channel, characterised by parameters $\xi = 1$ and $\xi=2$. The different figures represent different exemplary types of losses. We use these examples to build an understanding for when the loss channel and the photon subtraction commute. Note that if the losses act before the photon subtraction, we can simply replace $V$ with $V_{\xi}$ in (\ref{eq:wig}) and (\ref{eq:Ag}) to include the losses in the Gaussian state from which the photon is subsequently subtracted.\\

In Fig.~\ref{fig:NodeLossGraph}, we observe what happens when there is only a single lossy mode, being the one associated to the vertex where the photon was subtracted. We see that, independent of the value of $\xi$, only the local Wigner function for the vertex of subtraction is infected by the loss. The remainder of the state is completely unchanged. When we consider photon subtraction prior to the losses, this observation makes sense: %losses are a local physical process and therefore are bound to obey the no-signalling theorem. In other words, 
due to the no-signalling theorem we cannot alter parts of an entangled state by performing local operations (such as local losses) on different parts of the state. 

To understand the interplay of losses and photon subtraction, it is instructive to use (\ref{eq:D}) and write
\begin{equation}\label{eq:SpectTheorem}
e^{\xi D} = \sum_{j=1}^m e^{\xi \frac{\gamma_j}{2}} (P_{h_j}+P_{Jh_j}).
\end{equation}
In the case scenario of Fig.~\ref{fig:NodeLossGraph}, we can treat the problem in the basis of graph vertices and set $h_j = (0,\dots,0,1,0,\dots,0) \equiv e_j \in {\cal N}(\mathbb{R}^{2m})$, where the $1$ occurs on the $j$th position. Without loss of generality, we set $\gamma_j = \delta_{j1}$, with the subtraction mode $g = d_1$, the first vertex of the graph (which we henceforth refer to as ``vertex 1''). Hence, we straightforwardly see that $\tilde{g} = g$ in (\ref{eq:lossFinalOneSub}), which implies that losses and photon subtraction commute. This is counterintuitive, given that losses prior to photon subtraction affect the covariance matrix and reduce the entanglement between the mode of subtraction and the remaining modes. The spread of non-Gaussian features due to photon subtraction depends on these entanglement properties \cite{Walschaers-graphs-2018}, and, hence, it is far from clear how the reduced Wigner functions for the vertices 2, 3, and 4 can be unaffected by the losses (as is the case when the losses act after the photon subtraction). However, this can be understood from the conceptual approach in Fig.~\ref{fig:LossesMultimode}, where $O_2 = \mathds{1}$ in the present scenario.

In Fig.~\ref{fig:UniformLossGraph}, we show the scenario of uniform loss in all the modes. This implies that $D$ is a multiple of the unit matrix, and thus there is no preferred basis to describe the losses. To be consistent with the other cases, we set $D = \mathds{1}/2$. This choice makes it obvious that the losses and photon subtraction commute. 

As there is no preferred basis, we can again choose to treat the problem in the vertex basis to gain physical insight. We can interpret the losses as a combination of single-mode loss channels, which is essentially what we mathematically achieve through (\ref{eq:SpectTheorem}). In this sense, we just repeat the scenario of Fig.~\ref{fig:NodeLossGraph} for each mode. Note that, indeed, vertex 1 behaves exactly the same in Figs.~\ref{fig:NodeLossGraph} and \ref{fig:UniformLossGraph}. Invoking the no-signalling theorem again implies that the loss in one vertex cannot affect anything that happens in the other vertices. Because photon subtraction involves post-selection, it requires an action on all modes (the entire state is conditioned on the outcome of a local measurement). Hence, the argument based on no-signalling cannot be applied to mode-selective photon subtraction \cite{Walschaers-graphs-2018}. However, in this case, our argument of Section~\ref{sec:concept} shows that locally, each mode is only affected by the local loss, even when the photon subtraction is executed after the losses. As a consequence, we see a homogenous decay on the non-Gaussian features in each mode, as characterised by the minimal excess kurtosis (\ref{eq:kurt}).\\

There is no reason why the modes of the loss channel must coincide with graph's vertices, and, thus, we investigate a case of a non-local loss channel in Fig.~\ref{fig:Local-Comp}. As in Fig.~\ref{fig:NodeLossGraph}, we consider only a single-mode loss, but this time the mode is given by a balanced superposition of vertices 2, 3, and 4, i.e. $d = (e_2 + e_3 + e_4)/\sqrt{3}$ such that $d$ does not overlap with the subtraction mode $g$. This implies that $D = (P_d + P_{Jd})/2$, and from the spectral decomposition (\ref{eq:SpectTheorem}), we find that $g = \tilde{g}$ in (\ref{eq:lossFinalOneSub}). The losses and the photon subtraction, again, commute, and no-signalling tells us that vertex 1, where the photon is subtracted, remains unaffected by loss. The remaining vertices, however, display a different behaviour than what we saw before in Fig.~\ref{fig:UniformLossGraph}. %We observe that this non-local loss channel decreases the non-Gaussian features of the vertices 2 and 3 slightly faster than for the uniform losses of Fig.~\ref{fig:UniformLossGraph}, as indicated by the minimal excess kurtosis (\ref{eq:kurt}). On the other hand, the colour code indicates that vertex 4 remains more non-Gaussian for the same amount of loss as compared to Fig.~\ref{fig:UniformLossGraph}.\\

The most intriguing behaviour is expected to be found when, in addition, the losses act in a mode basis that has a non-trivial overlap with the mode of subtraction. This scenario is presented in Fig.~\ref{fig:ModeLossGraph}, where the losses act in a single mode, given by $d = (e_1 + e_4)/\sqrt{2}$. Again, this implies that we can set $D = (P_d + P_{Jd})/2$, and the spectral decomposition (\ref{eq:SpectTheorem}) dictates that $g \neq \tilde{g}$ in (\ref{eq:gTilde},\ref{eq:lossFinalOneSub}). This implies that we can interpret the photon subtracted state after losses as photon subtraction in mode $\tilde{g}$ from a Gaussian state with covariance matrix $V_{\xi}$, with
\begin{equation}
\tilde{g} = \frac{e^{\xi/2} +1}{\sqrt{2(1+e^{\xi})}}e_1 + \frac{e^{\xi/2} - 1}{\sqrt{2(1+e^{\xi})}}e_4.
\end{equation}
We clearly see that the mode $\tilde{g}$ changes with $\xi$, and that, indeed $\tilde{g} = g$ for $\xi=0$, whereas $\tilde{g} \approx d$ when $\xi$ is large. This implies that for weak losses, the state still looks similar to the original photon-subtracted state, whereas for strong losses, it seems as if the photon was subtracted from the mode $d$ in which the losses are taking place. As a consequence, the loss channel does not commute with the photon subtraction, as shown in Fig.~\ref{fig:ModeLossGraph}. In particular, we see that the no-signalling condition implies that vertices 2 and 3 remain unaffected when the losses occur after the photon subtraction. 

However, the scenario changes when the losses act on the Gaussian graph state prior to photon subtraction. As a most profound difference, we note at the bottom of Fig.~\ref{fig:ModeLossGraph} that the minimal excess kurtosis of vertices 2 and 3 is affected when the losses act prior to the photon subtraction, in strong contrast to the case where the photon subtraction is executed first. Furthermore, we note that the non-Gaussian features of vertex 4 vanish faster in the scenario where the losses act first. This difference is a direct consequence of the fact that we are now considering a case where $O_2 \neq \mathds{1}$ in the representation of Fig.~\ref{fig:LossesMultimode}.\\ %In particular, we find that $(V_{\xi} - \mathds{1})$ changes in a non-trivial way with $\xi$ as compared to the subtraction mode $g$, such that our argumentation in (\ref{eq:thisEq} - \ref{eq:thatEq}) no longer holds.\\

It should be emphasised that it is in practice important to understand whether or not losses commute with photon subtraction. In typical continuous-variable experiments, it is common practice to use Gaussian properties of the states, operations, and detectors when dealing with losses. In particular, when everything is Gaussian, one can use ideal models for optical elements and detectors and fully move the loss to the level of the quantum state. This is merely a convenient way of modelling, and it doesn't change the physics. However, because generally photon subtraction and losses do not commute, this approach will no longer hold. This is very relevant in experiments where the initial Gaussian state is characterised through homodyne measurement, and one obtains of covariance matrix $V$ for the state, which actually includes the detector losses. However, this covariance matrix $V$ can, in principle, not be used to model photon subtraction in the experiment, since part of the losses occur only at the stage of the measurement. This illustrates the importance of a good understanding of the place in the experiment where losses are occurring when dealing with photon subtracted states. We note that this is a way to even improve the agreement between theory and experiment in experiments such as \cite{Ra2019}.

%However, when (\ref{eq:SpectTheorem}) is inserted in eqs.~(\ref{eq:Vxi}) and (\ref{eq:Agfinal}), we can immediately understand that the restriction of $V_{\xi}$ and $A^{\xi}_{\tilde{g}}$ to all vertices but the one on which the losses act is independent of $\xi$ and unchanged through the loss channel.

\section{Conclusions and Outlook}

In summary, we have used the framework of open quantum systems and Lindblad equations to develop a model that describes losses from photon-subtracted quantum states of light. As a key result, we find in eq.~(\ref{eq:lossFinal}) that a loss channel maps an $n$-photon-subtracted state into a new $n$-photon subtracted state. In general, the modes of subtraction change due to the losses as described by eq.~(\ref{eq:lossFinal}). Furthermore, we show that the obtained results are equivalent to alternative models that describe losses in terms of a beam splitter that mixes the state with a certain amount of vacuum. 

The detrimental effect of the losses is illustrated in eqs.~(\ref{eq:wigloss}) and (\ref{eq:Agfinal}) on the level of the Wigner function of a single-photon-subtracted Gaussian state. These results lead us to inequality (\ref{eq:negcondnew}), that serves as a general condition for the existence of negative values of the Wigner function. In Fig.~\ref{fig:neg1mode}, the negativity of the Wigner function is probed for a single-mode example, which clearly shows that losses ultimately make the Wigner function positive.

In multimode scenarios, losses can be mode dependent. This can lead to a considerable change in the multimode structure of the non-Gaussian features in the state. To illustrate these multimode features, we consider photon-subtracted continuous variable graph states as an example in Figs.~\ref{fig:NodeLossGraph} - \ref{fig:Local-Comp}. In contrast to the single-mode scenario, we find that losses do not necessarily commute with photon subtraction. This effect is confirmed in the example shown in Fig.~\ref{fig:Local-Comp}, and emphasises the importance of knowing where losses occur in experimental setups.\\

This paper describes in very general terms how Gaussian losses act on photon-subtracted states. In this sense, these results are indispensable to interpret the details of state of the art multimode photon subtraction experiments \cite{ra_tomography_2017,Ra2019}.  Nevertheless, the outcomes of this study also offer interesting opportunities for the engineering of quantum states and measurements. Indeed, a thorough understanding of the losses in the system can help to identify the best modes to subtract a photon and let the non-Gaussian properties survive as long as possible. Furthermore, the understanding how the modal structure of the photon subtraction changes due to the losses can be used to develop an optimal homodyne measurement setup to extract the non-Gaussian features of the state. The homodyne detector itself is an important source of mode-dependent loss, which occurs after the photon subtraction. Our presented result shows that it is inaccurate to include these losses in the initial Gaussian state, even though this is common practice in most continuous-variable quantum optics experiments.

On a broader level, the results on graphs states in Figs.~\ref{fig:NodeLossGraph} - \ref{fig:Local-Comp} may also be relevant for quantum networking and quantum communication. In particular, one may consider to exploit mode dependent losses to transfer non-Gaussian features from one mode to another in a cleverly chosen mode basis.

\begin{acknowledgements} This work is supported by the French National Research Agency projects COMB and SPOCQ, and the European Union Grant QCUMbER (no. 665148). M.W. is funded through research fellowship WA 3969/2-1 from the German Research Foundation (DFG). Y.-S.R. acknowledges support from the European Commission through Marie Sk\l{}odowska-Curie actions (no. 708201) and support from the National Research Foundation of Korea (NRF) grant funded by the Korea government (MSIT) (No. NRF-2019R1C1C1005196). N.T. acknowledges financial support of the Institut Universitaire de France. 
\end{acknowledgements}

\appendix

\section{The interpretation of operators $a^{\dag}(Xf)$ and $a(Xf)$.}\label{app:creation}

Throughout this paper, we regularly use expression of the type $a^{\dag}(Xf)$ and $a(Xf)$ for $f \in {\cal N}(\mathbb{R}^{2m})$ and $X$ a $2m \times 2m$ matrix. These expressions may be confusing because in general $\norm{Xf} \neq 1$. Let us denote, for simplicity, that $Xf = \alpha \in \mathbb{R}^{2m}$. Hence, one must understand how to interpret $a^{\dag}(\alpha)$ (the interpretation of $a(\alpha)$ is analogous). 

Let us start by considering an arbitrary mode basis $\{u_1({\bf r}, t), \dots, u_m({\bf r},t)\}$ of our system. For every one of the modes, we have an associated creation operator $a^{\dag}_j$ that creates a photon in the $j$th mode. Moreover, to these modes, we associate a symplectic basis $\{e_1, \dots, e_m, Je_1, \dots J e_m\}$ of the phase space $\mathbb{R}^{2m}$. This basis is associated to the modes in the sense that 
$a^{\dag}(e_j) = a^{\dag}_j$, and applying (\ref{eq:crean}) implies that $a^{\dag}(Je_j) = ia^{\dag}_j$. We can now write 
\begin{equation}
\alpha = \sum_{j=1}^m [(e_j,\alpha)\mathds{1} + (Je_j, \alpha)J]\,e_j.
\end{equation}
We can than use the linearity of $a^{\dag}$ (which follows from the linearity of the quadrature operator $Q$) to find that
\begin{equation}
a^{\dag}(\alpha) = \sum_{j=1}^m [(e_j, \alpha) + i (Je_j, \alpha)] \, a^{\dag}_j,
\end{equation}
we can then insert that $\alpha = Xf,$ where $f \in {\cal N}(\mathbb{R}^{2m})$ can be associated to some mode, and obtain
\begin{equation}
a^{\dag}(Xf) = \sum_{j=1}^m [(e_j, Xf) + i (Je_j, X f)] \, a^{\dag}_j.
\end{equation}

An alternative way of understanding $a^{\dag}(Xf)$ is by defining a new mode, with associated with $\tilde{f} = Xf / \norm{Xf} \in {\cal N}(\mathbb{R}^{2m})$. One then finds that $a^{\dag}(Xf) = \norm{Xf} a^{\dag}(\tilde{f})$. When we compare $a^{\dag}(f)$ to $a^{\dag}(Xf)$, we note that the action of the matrix $X$ both changes the mode and rescales the operator. When we consider losses in the Heisenberg picture, it is common that $\norm{Xf} \leqslant 1$, such that we effectively see a decay in photon number, coherences, and correlations.

\section{Derivation of (\ref{eq:important})}\label{app:A}
Here we present two different derivations for the identity 
\begin{equation}\tag{\ref{eq:important}}\begin{split}
&a^{\dag}(g_n) \dots a^{\dag}(g_1) \Lambda_{\xi}(x) a(g_1)\dots a(g_n)\\
&=\Lambda_{\xi} [a^{\dag}(e^{\xi D}g_n) \dots a^{\dag}(e^{\xi D}g_1) x\,a(e^{\xi D}g_1)\dots a(e^{\xi D}g_n)],
\end{split}
\end{equation}
which lies at the heart of the derivation of the noise model. The first derivation uses the structure of the map $\Lambda_{\xi}$ as given by (\ref{eq:lambdaHeisSol}). The second approach that we sketch uses the structure of the master equation (\ref{eq:lindblad}) with Lindblad operators (\ref{eq:lindbladops}). 

\subsection{Derivation via the map $\Lambda_{\xi}$}

Let us first use (\ref{eq:lambdaHeisSol}) to show that
\begin{widetext}
\[\begin{split}\label{eq:longone}
&a^{\dag}(g_n) \dots a^{\dag}(g_1) \Lambda_{\xi} [a^{\dag}(f_1)\dots a^{\dag}(f_r)a(f_{r+1})\dots a(f_s)]  a(g_1)\dots a(g_n)\\
&=a^{\dag}(g_n) \dots a^{\dag}(g_1) a^{\dag}(e^{-\xi D}f_1)\dots a^{\dag}(e^{-\xi D}f_r)a(e^{-\xi D}f_{r+1})\dots a(e^{-\xi D}f_s) a(g_1)\dots a(g_n)\\
&=\Lambda_{\xi} [a^{\dag}(e^{\xi D}g_n) \dots a^{\dag}(e^{\xi D}g_1) a^{\dag}(f_1)\dots a^{\dag}(f_r)a(f_{r+1})\dots a(f_s)a(e^{\xi D}g_1)\dots a(e^{\xi D}g_n)],
\end{split}
\]
\end{widetext}
where we used that $e^{-\xi D}$ is invertible and has inverse $e^{\xi D}$. 

Because the creation and annihilation operators are generators of the algebra\footnote{Note, more accurately, they generate any operator in the Fock representation of the algebra.} of observables ${\cal A}$, we can approximate any observable $x \in {\cal A}$ by a polynomial of creation and annihilation operators. Through application of the canonical commutation relations, we can then cast all terms in this polynomial in normal order, to obtain the series expansion
\begin{equation}\label{eq:xExpansion}
x = \sum_{n_1, \dots, n_m = 0}^{\infty}\sum_{n'_1, \dots, n'_m = 0}^{\infty}  X^{n_1\dots n_m}_{n'_1 \dots n'_m}\, (a^{\dag}_1)^{n_1}\dots (a^{\dag}_m)^{n_m}(a_m)^{n'_1}\dots (a_1)^{n'_m},
\end{equation}
where $a^{\dag}_1, \dots, a^{\dag}_m$ and $a_1, \dots, a_m$ are creation and annihilation operators, respectively, for a randomly chosen mode basis, and $X^{n_1\dots n_m}_{n'_1 \dots n'_m} \in \mathbb{C}$ are the coefficients of the polynomial which represents $x\in {\cal A}$.
Hence, the result in (\ref{eq:lambdaHeisSol}) can be used to describe the full action of the channel on an arbitrary observable $x$. 

When we use the linearity of $\Lambda_{\xi}$, it follows that for all $x \in {\cal A}$
\begin{equation}\tag{\ref{eq:important}}\begin{split}
&a^{\dag}(g_n) \dots a^{\dag}(g_1) \Lambda_{\xi}(x) a(g_1)\dots a(g_n)\\
&=\Lambda_{\xi} [a^{\dag}(e^{\xi D}g_n) \dots a^{\dag}(e^{\xi D}g_1) x\,a(e^{\xi D}g_1)\dots a(e^{\xi D}g_n)].
\end{split}
\end{equation}

\subsection{Derivation via the master equation}
One may argue that some of the steps in the previous derivation lack elegance, even though it the derivation is relatively easy. A slightly more appealing alternative can be obtained by considering the master equation (\ref{eq:lindblad}) with Lindblad operators (\ref{eq:lindbladops}):
\begin{equation}\label{eq:lindbladApp}
\frac{\rm d}{{\rm d} \xi} \Lambda_{\xi}(x) = {\cal L}(x), \quad {x \in {\cal A}},
\end{equation}
with
\begin{equation}
{\cal L}(x) = \sum_j \gamma_j\left[ a^{\dag}(h_j) x a(h_j) -\frac{1}{2} \{a^{\dag}(h_j)a(h_j) , x\}\right].
\end{equation}
We can than use the canonical commutation relation
\begin{equation}
[a(f_1),a^{\dag}(f_2)] = (f_1,f_2) + i (f_1, J f_2),
\end{equation}
to obtain that
\begin{align}\label{eq:lindidentity}
a^{\dag}(g){\cal L}(x)a(g) = &{\cal L}[a^{\dag}(g) x \,a(g)] \\&+ a^{\dag}(Dg) x\, a(g) + a^{\dag}(g) x \,a(Dg).\nonumber
\end{align}
We can formally solve the master equation (\ref{eq:lindbladApp}) as $\Lambda_{\xi} = \exp(\xi {\cal L})$, where the exponential of the super-operator is defined in terms of a series expansion. Thus, we obtain that
\begin{equation}
a^{\dag}(g)\Lambda_{\xi}(x)a(g) = \sum_{n=0}^{\infty} \frac{\xi^n}{n!} a^{\dag}(g)\underbrace{{\cal L}\circ \dots \circ {\cal L}}_{\times n}(x)a^{\dag}(g),
\end{equation}
where we can repeatedly apply (\ref{eq:lindidentity}) and regroup the terms.  After some straightforward calculations, we find that for all $x \in {\cal A}$ \begin{equation}\tag{\ref{eq:important}}\begin{split}
&a^{\dag}(g_n) \dots a^{\dag}(g_1) \Lambda_{\xi}(x) a(g_1)\dots a(g_n)\\
&=\Lambda_{\xi} [a^{\dag}(e^{\xi D}g_n) \dots a^{\dag}(e^{\xi D}g_1) x\,a(e^{\xi D}g_1)\dots a(e^{\xi D}g_n)],
\end{split}
\end{equation}
as expected.

\bibliography{notes_losses.bib}

\end{document}